\documentclass[preprint,12pt]{elsarticle}

\usepackage{amssymb}

\usepackage{amsmath,amsfonts}
\usepackage{graphicx}
\usepackage{float}
\usepackage{subfig}
\usepackage{verbatim} %comment
\usepackage{color}

\usepackage{booktabs}
\usepackage{tablefootnote}
\usepackage{threeparttable}
\usepackage{makecell}

\journal{}

\begin{document}

\begin{frontmatter}



\title{Persistent Homology-Driven Optimization of Effective Relative Density Range for Triply Periodic Minimal Surface}

\author[1]{Depeng Gao}
\author[1]{Yuanzhi Zhang}
\author[1]{Hongwei Lin\corref{cor1}}

\affiliation[1]{organization={School of Mathematical Sciences, Zhejiang University},
            addressline={No.866, Yuhangtang Rd}, 
            city={Hangzhou},
            postcode={310058}, 
            state={Zhejiang Provence},
            country={China}}

\cortext[cor1]{Corresponding author. E-mail address: hwlin@zju.edu.cn (H. Lin).}


\begin{abstract}
    Triply periodic minimal surfaces (TPMSs) play a vital role in the design of porous structures, with applications in bone tissue engineering, chemical engineering, and the creation of lightweight models. 
    However, fabrication of TPMSs via additive manufacturing is feasible only within a specific range of relative densities, termed the effective relative density range (EDR), outside of which TPMSs exhibit unmanufacturable features. 
    In this study, the persistent homology is applied to theoretically calculate and extend the EDRs of TPMSs. 
    The TPMSs with extended EDRs are referred to as extended TPMSs. 
    To achieve this, TPMSs are converted into implicit B-spline representation through fitting. 
    By analyzing the symmetry of TPMSs, a partial fitting method is utilized to preserve the symmetry and enhance fitting precision.  
    A topological objective function is modeled based on the understanding of topological features, resulting in extended TPMSs that possess extended EDRs while maintaining a high degree of similarity to the original TPMSs.  
    Experimental validation confirms the effectiveness of the approach in extending the EDRs of TPMSs. 
    Furthermore, the extended TPMSs demonstrate superior performance in porous model design and topology optimization compared to their original counterparts. 
    The extended TPMSs with increased EDRs hold promise for replacing traditional TPMSs in applications that require porous structures with varying densities.
 
\end{abstract}

\begin{graphicalabstract}
\includegraphics[width=0.99\textwidth]{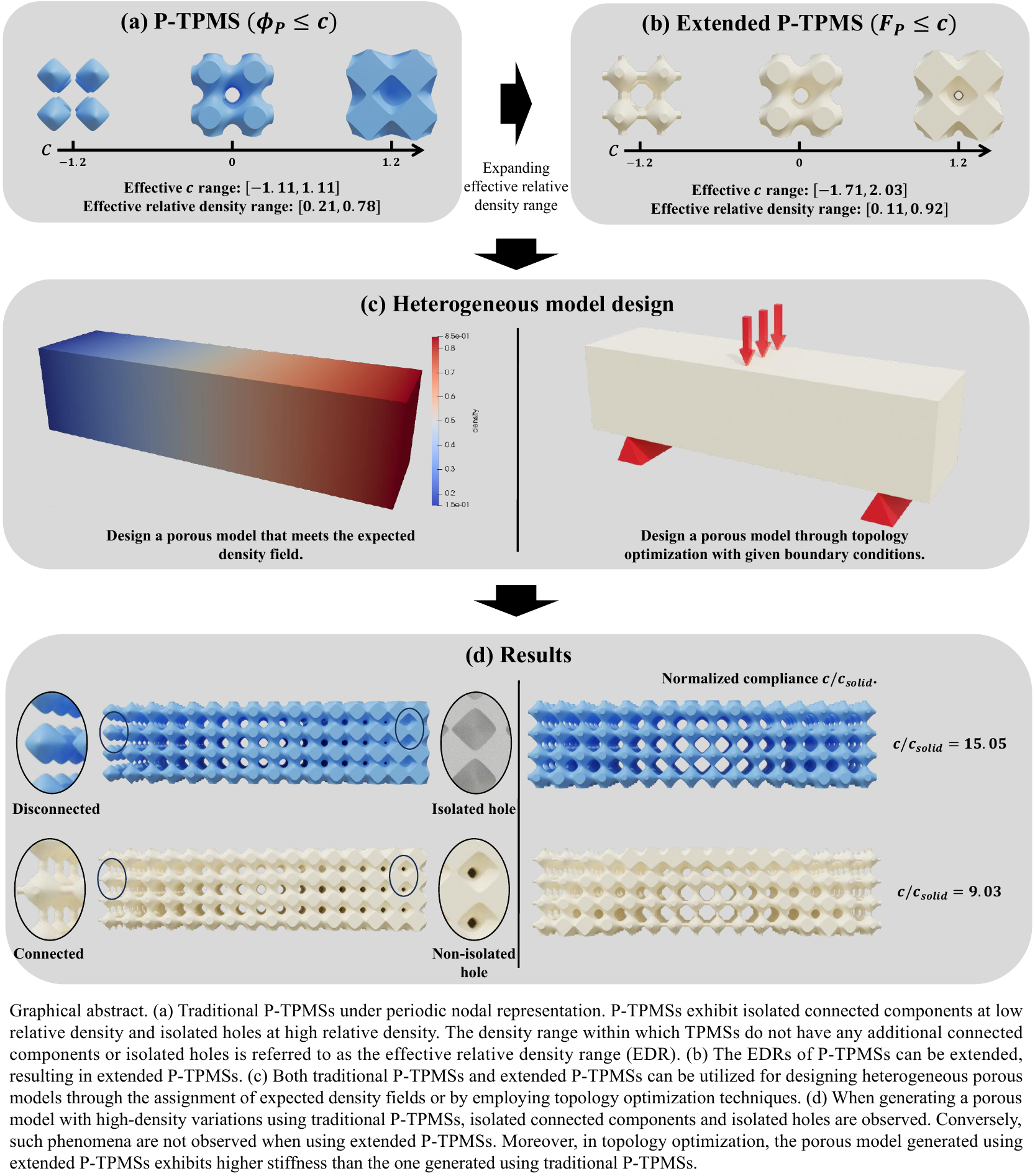}
\end{graphicalabstract}

\begin{highlights}
    \item[1.] This study proposes a method for calculating and analyzing effective threshold ranges and effective relative density ranges of TPMSs from a topological perspective using persistent homology.
    \item[2.] New representations of TPMSs using B-spline functions are introduced, providing greater controllability compared to traditional TPMSs.
    \item[3.] A new topological objective function is formulated to extend the effective relative density ranges of TPMSs while maintaining the similarity to the initial structures.
    \item[4.] The experiments show that the extended TPMS demonstrates better performance in high stiffness model design compared to the original TPMS.  
\end{highlights}


\end{frontmatter}


\section{Introduction} 
\label{sec: Introduction}
The advent of additive manufacturing (AM) has revolutionized the production of complex structures, enabling the creation of designs that were previously unattainable with traditional subtractive manufacturing methods. 
Porous structures have gained significant attention in the AM community due to their potential for achieving high strength-to-weight ratios and specific surface area, factors that are critical for applications in aerospace, medical, and chemical engineering.
Although chemical methods like foaming have been commonly used for fabricating porous structures, AM offers the advantage of precisely controlling the pore features of the final model through combining computer-aided design methods~\cite{feng2018review}. 
Therefore, designing complex porous models that exhibit desired physical properties, such as relative density and pore size, while ensuring their manufacturability is a meaningful issue. 
Among the various structures used for porous models, Triply Periodic Minimal Surfaces (TPMSs) have gained prominence due to their remarkable biological and mechanical performances. 
TPMSs possess smooth structures, self-supporting characteristics, and a small storage footprint, making them increasingly popular in the design of modern porous models.

\begin{figure} [h] 
	\centering
\includegraphics[width=0.8\textwidth]{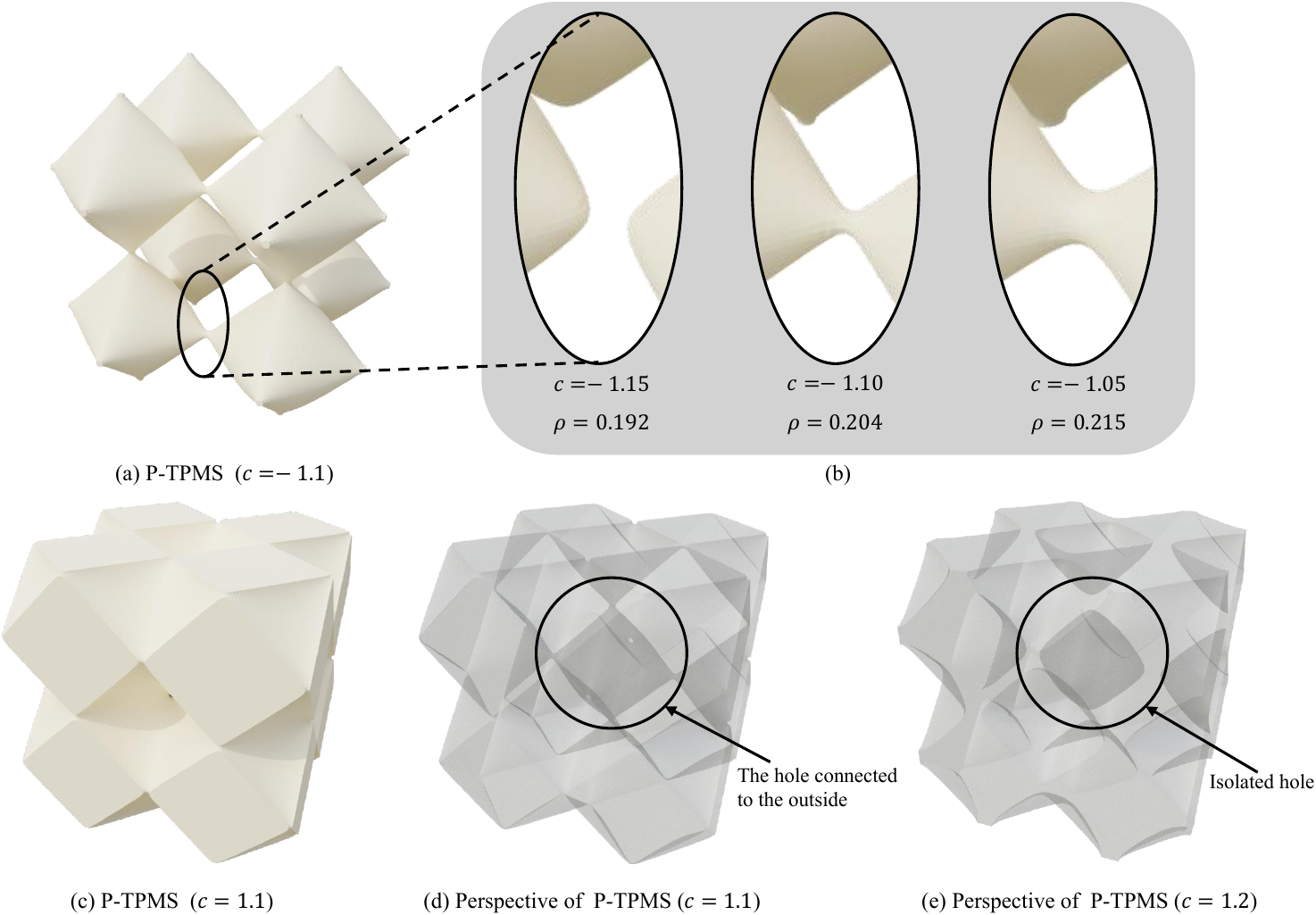}
	\caption{Rod P-TPMS structures ($\phi_{P}\leq c$) under different threshold $c$. (a) Rod P-TPMS with $c=-1.1$. (b) The porous structure has numerous connected components when $c=-1.15$. (c) Rod P-TPMS with $c=1.1$. (d) The porous structure has no isolated holes when $c=1.1$. (e) The porous structure has an isolated hole when $c=1.2$.}
	\label{fig: P_disadvantage_unit} 
\end{figure}

Although TPMSs have found wide application, some types have limited ability to achieve a continuous variation in relative density from 0 to 1, which negatively impacts their performance in various applications. 
The TPMSs represented by implicit functions $\phi = c$ divide the space into two separate parts, and the part $\{(x,y,z)~|~\phi(x,y,z) \leq c\}$ can be used to represent a solid structure~\cite{hu2021heterogeneous,gao2023free,feng2022stiffness}. 
Figure~\ref{fig: P_disadvantage_unit} shows that the relative density of the solid structure increases with higher values of $c$.  
For small values of $c$ (Figure~\ref{fig: P_disadvantage_unit}(b)), the porous structure consists of multiple connected components. 
Conversely, for large values of $c$ (see Figure~\ref{fig: P_disadvantage_unit}(e)), isolated holes appear within the porous structure. 
The isolated connected components lack support, and the isolated holes impede the discharge of the waste liquid during manufacturing, making the porous model unmanufacturable. 
The range of $c$ values without additional connected components and isolated holes is defined as the \textbf{effective threshold range (ETR)}. 
Since the relative density is monotonically related to the threshold, the corresponding range of relative density values for the ETR is called the \textbf{effective relative density range (EDR)}. 

\begin{figure} [h] 
	\centering
	\subfloat[]{
		\includegraphics[width=0.48\textwidth]{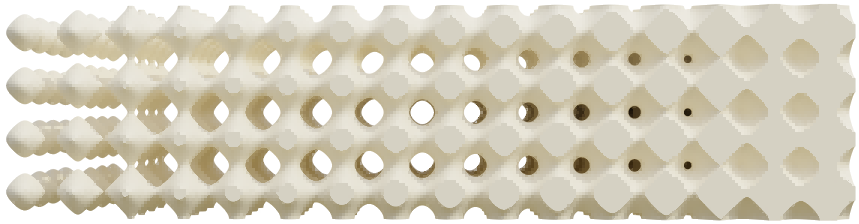}} 
    \subfloat[]{
		\includegraphics[width=0.48\textwidth]{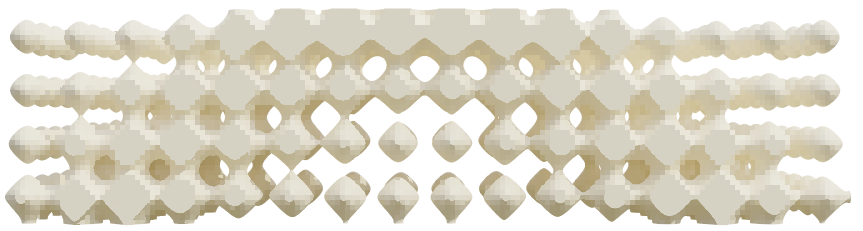}}  
	\caption{Heterogeneous P-TPMS models. (a) From left to right, the relative density of the model gradually increases from 0.1 to 0.9. (b) A porous model obtained by setting the target volume ratio to 0.2 in topology optimization.}
	\label{fig: P_disadvantage} 
\end{figure}

Porous structures with a wider EDR provide a broader range for designing. 
    Figure~\ref{fig: P_disadvantage}(a) illustrates the design of a porous model using P-TPMS, showing a continuous density range from 0.1 to 0.9. 
    However, this range exceeds the EDR of P-TPMS, resulting in numerous connected components on the left side and isolated holes on the right side of the model. 
    Consequently, the manufacturing of this porous model is not feasible. 
    Figure~\ref{fig: P_disadvantage}(b) demonstrates the use of P-TPMS in topology optimization to generate a porous model with a volume constraint of 0.2. 
    However, since the volume constraint of 0.2 falls beyond the EDR of P-TPMS, it becomes impossible to obtain an optimized model that can be manufactured. 
    In conclusion, using a porous with broader EDR enables the design of a porous model with a more extensive range of density variation.  
    Furthermore, enlarging the EDR of the porous can extend the solution space in the topology optimization problem, resulting in a solution that are closer to the optimum~\cite{liu2023multiscale}.

Some studies have noticed the above phenomenon of TPMSs and limited the threshold range of TPMSs in the design of porous models~\cite{hu2021heterogeneous,xu2023topology,li2018optimal}. 
    Xu et al.~\cite{xu2023topology} discussed and presented the ETRs and EDRs for different TPMSs. 
    However, they did not provide a theoretical calculation method for the ETRs and EDRs. 
    Li et al.~\cite{li2018optimal} introduced a penalty function to the TPMS of G-type to remove additional connected components and isolated holes in situations of low and high relative density, respectively. 
    However, this method cannot directly extend to other types of TPMSs. 
    Therefore, building upon the above studies, this study proposes a theoretical calculation method for EDRs and ETRs from a topological perspective using persistent homology. 
    Additionally, the EDRs of TPMSs are extended while maintaining the underlying structures of the initial range, thereby improving their performance in model design. 
    Firstly, the EDRs and ETRs of TPMSs are analyzed and calculated using persistent homology. 
    Next, TPMSs under periodic nodal representations are converted to implicit B-spline function representation through fitting.
    Subsequently, the symmetries of TPMSs are analyzed, and a partial fitting method is employed to preserve cubic symmetry and improve fitting precision. 
    Lastly, a novel topological objective function is formulated to extend the EDRs while maintaining the similarity of the optimized structures to the initial structures, resulting in extended TPMSs. 
    The main contributions of this study are as follows:
    \begin{itemize}
    \item[1.] This study proposes a method for calculating and analyzing ETRs and EDRs of TPMSs from a topological perspective using persistent homology.
    \item[2.] New representations of TPMSs using B-spline functions are introduced, providing greater controllability compared to traditional TPMSs.
    \item[3.] A new topological objective function is formulated to extend the EDRs of TPMSs while maintaining the similarity to the initial structures.
    \item[4.] The experiments show that the extended TPMS demonstrates better performance in high stiffness model design compared to the original TPMS.  
    \end{itemize}

The structure of this paper is organized as follows: 
Section~\ref{subsec: Related work} reviews related works on the heterogeneous porous model design.
    Section~\ref{sec: Preliminaries} provides an introduction to B-spline functions, persistent homology, and TPMSs.  
    Next, Section~\ref{sec: Method} explains the method for generating the extended TPMSs.
    Section~\ref{sec: Implementation and discussion} presents experimental results that demonstrate the effectiveness of the method. 
    The study concludes with a summary in Section~\ref{sec: Conclusion}.

\subsection{Related work}
\label{subsec: Related work}
The mechanical and biological performance of a porous model is significantly influenced by its density distribution, highlighting the importance of controlling this distribution. 
    A porous model with varying relative densities and morphologies of pores at different spatial locations is referred to as a heterogeneous porous model.

One method to design heterogeneous models involves controlling the threshold $c$ in the implicit equations $\phi_{TPMS}=c$ of TPMSs.
    Several studies have replaced the constant threshold $c$ with a spatially varying threshold field to achieve heterogeneous porous structures~\cite{yoo2012heterogeneous,feng2018porous,feng2019sandwich}. 
    Feng et al. investigated the relationship between the threshold and anisotropy~\cite{feng2021isotropic}. 
    Hu et al. designed the threshold field guided by the density distribution of free-form models~\cite{hu2021heterogeneous}. 
    Another approach involves designing porous structures with different densities and using interpolation methods to smoothly transition and stitch them together to form a model~\cite{yang2014multi,yoo2015advanced,li2021simple}. 
    Although these two methods can effectively generate heterogeneous porous models, the density distribution of the designed models always mimics natural structures. 

The topology optimization method provides theoretical guidance for designing the density distribution of heterogeneous porous models. 
    The aim of topology optimization is to find the optimal material distribution that achieves the best performance under given boundary conditions. 
    Using the homogenization method~\cite{pinho2009asymptotic} allows for deriving the mathematical relationship between density and the homogenized elasticity tensor of TPMSs, enabling the direct application of TPMSs in topology optimization models~\cite{pinho2009asymptotic,feng2022stiffness,ozdemir2023novel,shi2021design}. 
    Li et al. used TPMSs to optimize the thermal and mechanical compliance of the porous model, which significantly improved its performance~\cite{li2019design}. 
    Hu et al. proposed an equivalent method to predict the elastic modulus of TPMSs and established a topology optimization model based on isogeometric analysis to enhance the stiffness of free-form models~\cite{hu2023isogeometric}. 
    Montemurro et al. introduced a numerical method to predict the equivalent thermal conductivity tensor of TPMSs and subsequently integrated it with the Solid Isotropic Material with Penalization method to improve the thermal conduction efficiency of the porous model~\cite{montemurro2022thermal}.

In the aforementioned methods for designing heterogeneous models, the density field is crucial in controlling the performance of the model. 
    Given a basic porous structure as the microscopic unit, designing a porous model is largely equivalent to designing the density field. 
    A porous structure with a larger Effective Design Range (EDR) provides a larger design domain for the density field, thus offering a greater range for designing models to achieve better performance.

\section{Preliminaries}
\label{sec: Preliminaries}
\begin{comment}
\begin{equation}
   \begin{aligned}
      B_{i,0}(u)=& \left\{
         \begin{aligned}
            1, \quad &t_i \leq u < t_{i+1}, \\
            0, \quad &otherwise 
         \end{aligned}
         \right.\\
       B_{i,p}(u)=&\frac{u-t_i}{t_{i+p}-t_i}B_{i,p-1}(u)+ \\ &\frac{t_{i+p+1}-u}{t_{i+p+1}-t_{i+1}}B_{i+1,p-1}(u).
   \end{aligned} 
\end{equation}
\end{comment}
\subsection{Trivariate B-spline function}
Let $\mathbf{T}=\{t_0,t_1,\ldots,t_{m-1}\}$ be a non-decreasing sequence of real values, and it is referred to as the \textbf{knot vector}. 
    The $i$-th B-spline basis $B_{i,p}(u)$ of degree $p$ is defined based on $\mathbf{T}$ using the deBoor-Cox formula~\cite{piegl1996nurbs}.

\textbf{A B-spline function} of degree $p$ is defined by:
\begin{equation}
   C(u)=\sum_{i=0}^{n-1}B_{i,p}C_i , \quad 0\leq u \leq 1,
\end{equation}
where $C_i$ is the $i$-th control coefficient, a real value, and $n$ is the number of control coefficients. 

Similarly, \textbf{a trivariate B-spline function} of degree $(p_u,p_v,p_w)$ can be defined as a tensor product using the following expression:
\begin{equation}
    \begin{aligned}
        &C(u,v,w) = \sum_{i=0}^{n_u-1}\sum_{j=0}^{n_v-1}\sum_{k=0}^{n_w-1}R_{ijk}(u,v,w)C_{ijk}. \\
        & R_{ijk}(u,v,w) = B_{i,p_u}(u)B_{j,p_v}(v)B_{k,p_w}(w),
    \end{aligned}
\end{equation}
where $R_{ijk}$ is referred to as the blending basis function.

\subsection{Topology-aware optimization}
\label{subsec:Topology-aware optimization}
Persistent homology~\cite{edelsbrunner2002topological,chazal2021introduction} is a power tool for inferring topological features of different scales from a discrete space that is filtered by a real-valued function. In this study, a porous structure is represented by an implicit function, which serves as the real-valued function to induce a sublevel filtration. Additionally, the space is discretized into a cubical complex to optimize the topological features of the porous structure.

\begin{figure} [h] 
	\centering
	\subfloat[]{
		\includegraphics[width=0.23\textwidth]{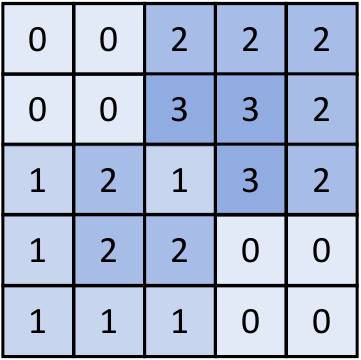}} 
	\subfloat[]{
		\includegraphics[width=0.24\textwidth]{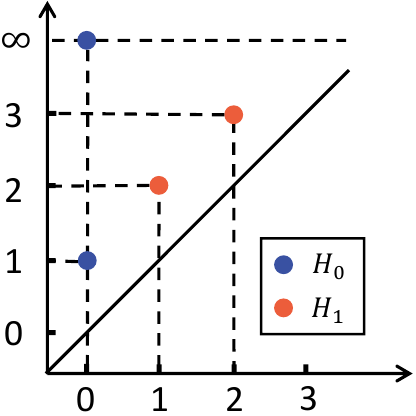}}
	\subfloat[]{ 
		\includegraphics[width=0.5\textwidth]{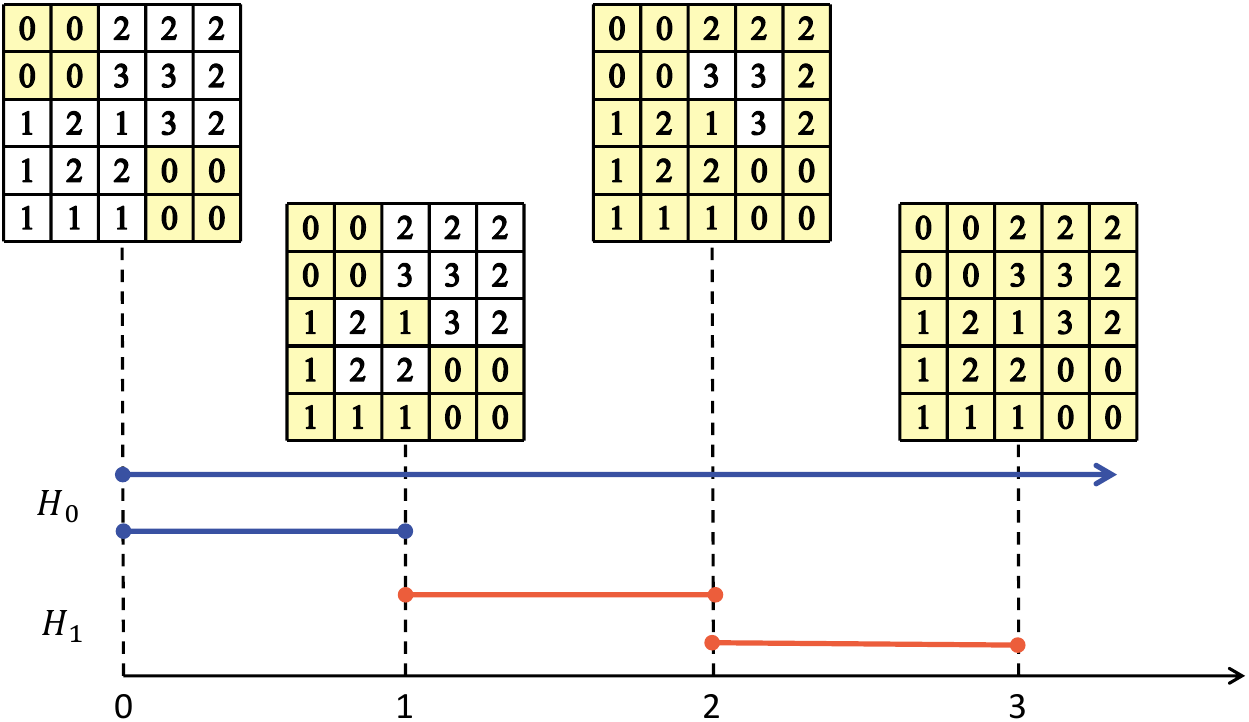}}
	\caption{Explanation of persistent homology. (a) A cubical complex with the function defined on it. (b) A persistence diagram records the persistent pairs. (c) A persistent barcode and cubical complexes (marked in yellow) in the filtration. At time 0, there are two connected components (0-dimensional topological features) in the corresponding underlying space. 
    At time 1, the two connected components merge into one, and one of them disappears. 
    Therefore, it corresponds to the point $(0,1)$ in the persistence diagram and the second line segment in the persistence barcode. 
    At time 1, a loop appears in the underlying space (1-dimensional topological feature), which disappears at time 2, corresponding to the point $(1,2)$ in the persistence diagram. 
    Because there is always one connected component in the underlying space, the point $(0,\infty)$ exists in the persistence diagram.}
	\label{fig:persistent homology} 
\end{figure}

$k$-cubes ($\kappa_k$) are the fundamental elements defined as the Cartesian product of $k$ unit intervals in Euclidean space, such as $0$-cubes (vertices), $1$-cubes (line segments), and $2$-cubes (squares). A cubical complex $\mathcal{K}$ is a collection of $k$-cubes, where the intersection of any two cubes is either empty or a common face of both. The union of all cubes in $\mathcal{K}$ forms its underlying space $\mathcal{S}$. Computational homology~\cite{kaczynski2004computational} has proposed methods to compute the topological invariants of a cubical complex $\mathcal{K}$, known as the homology group. The $d$-th homology group $H_d(\mathcal{K})$ encodes topological information of dimension $d$. The rank of $H_d(\mathcal{K})$, called the $k$-Betti number $\beta_d(\mathcal{K})$ indicates the number of homological classes in the k-dimension. From an intuitive but not rigorous perspective, the $\beta_0(\mathcal{K})$, $\beta_1(\mathcal{K})$, and $\beta_2(\mathcal{K})$ represent the number of connected components, loops, and cavities in the underlying space, respectively. 

Let $f$ be a real-valued function defined on $\mathcal{S}$. The value of a $k$-cube is the maximal value among all its vertices ($0$-cube), that is defined as:
\begin{equation}
    \label{eq: function on simplex}
    f(\kappa_k) = \max_{\kappa_0\in \kappa_k}f(\kappa_0).
\end{equation}
The cubical complex $\mathcal{K}_f^t$ is induced by $\mathcal{K}_f^t = \{\kappa \in \mathcal{K}~|~f(\kappa)\leq t\}$. The sublevel filtration $\mathcal{K}_f$ is defined as the collection of cubical complexes $\mathcal{K}_f^t$ for all $t$ in $\mathbb{R}$. It is a nested sequence of cubical complexes that satisfy $\mathcal{K}_f^{t_i} \subseteq  \mathcal{K}_f^{t_j}$ for all $t_i < t_j$, where $t_i$, $t_j \in \mathbb{R}$. Since the number of cubes is finite in practical applications, the filtration consists of a finite number of complexes. Assuming a topological feature (homological class) appears in complex $\mathcal{K}_f^b$ and disappears in $\mathcal{K}_f^d$, persistent homology tracks the "birth" and "death" of this topological feature and pairs them as a persistent pair $(b,d)$. All the persistent pairs in $d$-dimensional space are encoded as a collection of intervals called $d$-dimensional persistence barcodes (see Figure~\ref{fig:persistent homology}(c)). Alternatively, they can be represented as a multiset of points in $\mathbb{R}^2$ called $d$-dimensional persistence diagram (PD) (see Figure~\ref{fig:persistent homology}(b)).

Each topological feature is associated with a simplex that generates it and another simplex that terminates it. The topological inverse mapping proposed in~\cite{poulenard2018topological} realizes the mapping from the persistent pairs to the simplexes. The topological inverse mapping $\pi_f$ defines the corresponding simplex pair for a given persistent pair $(b,d)$ on PD as follows:
\begin{equation}
    (\sigma,\tau) = \pi_f(b,d), \quad \sigma,\tau\in\mathcal{K}.
\end{equation}
Therefore, the persistent pair $(b,d)$ can be written as $(f(\sigma),f(\tau))$. 
    Furthermore, from Equation~\ref{eq: function on simplex}, there exists a 0-simplex $\kappa_0$ such that $f(\sigma) = f(\kappa_0)$. 
    Therefore, a persistent pair can be further mapped to a pair of 0-simplexes:
    \begin{equation}
        \label{eq: map persistent pair to 0-simplexes}
        (\kappa_0^b,\kappa_0^d) = \tilde{\pi}_f(b,d). 
    \end{equation}

Assuming that the topological objective function $L_{top}$ is defined on the persistence diagram, and $f$ defined by the parameter family $\mathbf{\alpha}$, the partial derivative $\partial L_{top}/\partial \alpha $ can be calculated through the inverse mapping $\tilde{\pi}_f$. As mentioned in prior work by~\cite{gao2022connectivity,bruel2020topology,dong2022topology}, the real-valued function $f$ can be optimized by the objective function. However, it is essential to establish guidelines for defining a rational topological objective function and identifying a significant function $f$ to induce filtration.

\subsection{Triply periodic minimal surface}
\label{subsec: Triply periodic minimal surface}
TPMSs are algebraic surfaces represented by implicit equations. 
    Typically, TPMSs are represented using periodic nodal surfaces defined by Fourier series~\cite{hu2023isogeometric}:

\begin{equation}
    \label{eq: tpms}
    \Phi(\mathbf{r}) = \sum_{\mathbf{k}}F(\mathbf{k})\cos{2\pi \mathbf{k}\cdot\mathbf{r}-\alpha(\mathbf{k})}=0,
\end{equation}
where $\mathbf{k}$ represents the reciprocal lattice vectors, $\alpha(\mathbf{k})$ denotes the phase shift, and $F(\mathbf{k})$ is an amplitude associated with a given $\mathbf{k}$-vector. 
    High frequency items in Equation~\ref{eq: tpms} are truncated to generate commonly used formulas in porous model design. 
    The nodal approximations of several TPMSs presented in~\cite{hu2023isogeometric} are listed in Table~\ref{tab: TPMS}. 
    
Table~\ref{tab: TPMS} reveals that when $\omega_x=\omega_y=\omega_z=1$, TPMSs exhibit a periodicity of $2\pi$ in all three directions. 
    The TPMS structure within the domain $[0,2\pi]^3$ is referred to as a \textbf{complete unit}, while the structure within the range $[0,\pi]^3$ is called a \textbf{half unit}. 
    For convenience in discussions, the periodicity of all TPMSs in all three directions is set to $2\pi$ in the subsequent sections.

To manufacture TPMSs, solid TPMSs can be represented by the following equations using the implicit function $\phi_{TPMS}$~\cite{hu2021heterogeneous}:
\begin{itemize}
    \item \textbf{Rod} type: $\phi_{TPMS}\leq c$,
    \item \textbf{Pore} type: $\phi_{TPMS}\geq c$,
    \item \textbf{Sheet} type: $-c \leq \phi_{TPMS} \leq c$,
\end{itemize}
where $c$ is referred to as the threshold. 
    The threshold $c$ can be replaced with a function $c(u,v,w)$ to create a heterogeneous porous structure with various density distribution within the space. 

\begin{table}[]
\caption{Nodal approximations of TPMSs presented in~\cite{hu2023isogeometric}.}
\centering
\begin{threeparttable}
\resizebox{0.98\textwidth}{!}{
\begin{tabular}{cc}
\hline
Type & Nodal approximations \\ \hline
P    & $\phi_P(x,y,z)=\left[\cos (\omega_x x) + \cos (\omega_y y)+\cos (\omega_z z)\right]/0.9 = C$        \\
D    & $\phi_D(x,y,z)=\left[\cos (\omega_x x) \cos (\omega_y y) \cos (\omega_z z) - \sin (\omega_x x) \sin (\omega_y y) \sin (\omega_z z)\right]/0.6 = C$               \\
G    & $\phi_G (x,y,z) = \left[\sin (\omega_x x) \cos (\omega_y y) + \sin (\omega_y y) \cos (\omega_z z) + \sin (\omega_z z) \cos (\omega_x x)\right]/0.9 = C$                     \\
I-WP & \makecell{$\phi_{I-WP} (x,y,z) =\{ 2[ \cos (\omega_x x) \cos (\omega_y y) + \cos (\omega_y y) \cos (\omega_z z) + \cos (\omega_z z) \cos (\omega_x x) ]$ \\ $-  [ \cos (2\omega_x x) + \cos (2\omega_y y) + \cos (2\omega_z z) ]\}/2.5 = C$}                     \\ \hline
\end{tabular}}
\end{threeparttable}
\label{tab: TPMS}
\end{table}

\section{Expansion and calculation of EDRs in TPMSs}
\label{sec: Method}
This section introduces the method for calculating and extending the EDRs of TPMSs. 
    In Subsection~\ref{subsec: Topological Analysis of TPMSs}, persistent homology is used to analyze and calculate the EDRs and ETRs of TPMSs. 
    Subsection~\ref{subsec: Parametric representation of TPMSs} introduces the conversion of TPMSs into an implicit B-spline representation. 
    To enhance fitting accuracy and preserve the cubic symmetry of TPMSs, Subsection~\ref{subsec: Preservation of symmetry} improves the fitting method described in Subsection~\ref{subsec: Parametric representation of TPMSs}. 
    Subsection~\ref{subsec: Method for expanding the effective threshold range} proposes the topological objective function for extending the EDRs of TPMSs, resulting in extended TPMSs.

\subsection{Topological Analysis of TPMSs}
\label{subsec: Topological Analysis of TPMSs}
In this subsection, the P-TPMS of Rod-type $\{\phi_P\leq c\}$ is taken as an example to analyze the impact of the threshold on its topological structures. 
    This analysis method is also applicable to Pore-type and Sheet-type TPMSs.

\begin{figure}[h]
    \centering
    \includegraphics[width=0.6\textwidth]{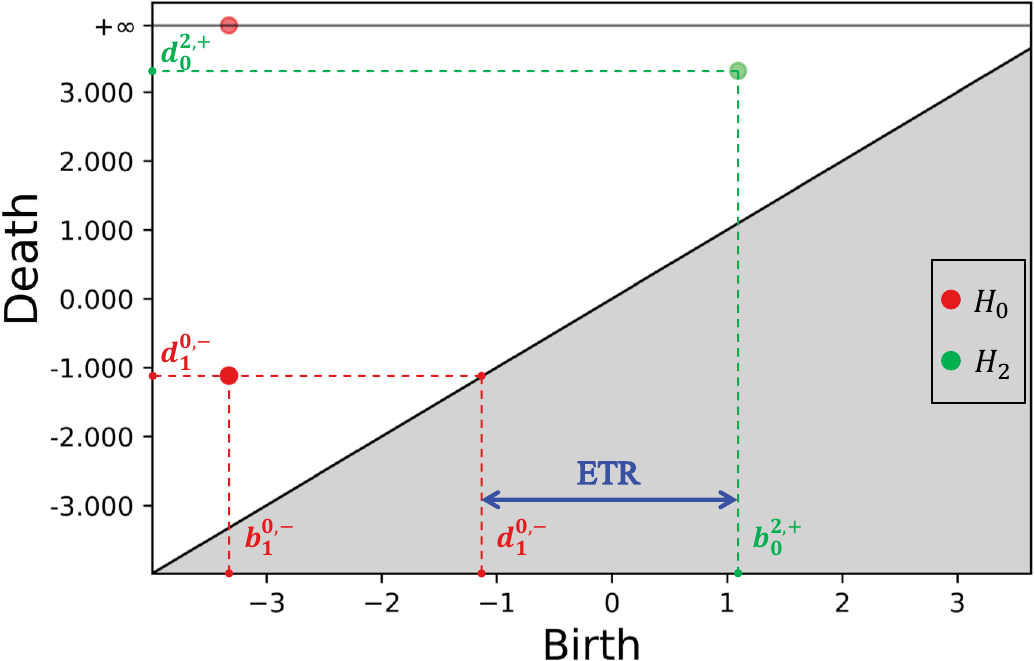}
    \caption{0-dimensional and 2-dimensional persistence diagrams of the Rod type P-TPMS. }
    \label{fig:PD of P-TPMS}
\end{figure}

As mentioned in Section~\ref{subsec:Topology-aware optimization}, when given a real-valued function $f$, the topological variation of the set $\{(u,v,w)|f(u,v,w)\leq t\}$ can be tracked using persistent homology. 
    In this case, the function $\phi_P$ is used as a substitute for the function $f$ to induce a filtration. 
    The "birth" and "death" of 0-dimensional and 2-dimensional topological features are then recorded in a persistence diagram, as shown in Figure~\ref{fig:PD of P-TPMS}.

Given a set of k-dimensional persistent pairs$\{(b_i^k,d_i^k)\}_{i=0}^{M_k-1}$, it can be reordered based on different rules:
\begin{itemize}
    \item Descending order based on the death time: $\{(b_i^{k,-},d_i^{k,-})\}_{i=0}^{M_k-1}$.
    \item Ascending order based on the birth time: $ \{(b_i^{k,+},d_i^{k,+})\}_{i=0}^{M_k-1}$.
\end{itemize}
    As illustrated in Figure~\ref{fig:PD of P-TPMS}, there are several persistent pairs that overlap at $(b_1^{0,-},d_1^{0,-})$, which correspond to the connected components as shown in Figure~\ref{fig: P_disadvantage_unit}(a). 
    All the 0-dimensional topological features excepted one are died when $c=d_1^{0,-}$.  
    Therefore, there is only one connected component in the Rod P-TPMS when $c\in [d_1^{0,-},+\infty)$. 
    Similarly, isolated holes appear when $c\in [b_0^{2,+},d_0^{2,+})$, and there are no isolated holes when $c \in (-\infty,b_0^{2,+})$. 
   
As the threshold $c$ increases, more material appears in the Rod P-TPMS. 
    When $c$ is small, there is only a small volume of material, making it more prone to isolated connected components. 
    When $c$ is large, due to the presence of a large volume of material, isolated holes will form. 
    For TPMSs, it is often true that $d_1^{0,-} < b_0^{2,+}$, which means that there is always a threshold range where the porous structure has no isolated holes and only one connected component.
    Notably, although there may still be certain threshold ranges where $c < d_1^{0,-}$ or $c\geq b_0^{2,+}$, and isolated holes and additional connected components do not exist in the porous structure, the porous structure at these moments is either void or approximates a solid cube.    
    Therefore, in the following discussion, only the interval $[d_1^{0,-}, b_0^{2,+})$ is considered as the ETR of the Rod TPMS. 
    The relative density of the Rod TPMS increases with the growth of the threshold $c$. 
    The EDR corresponding to the ETR is denoted by $[\rho_{min},\rho_{max}]$.

The same discussion is applicable to the Pore type and Sheet type TPMSs using the following transformations:
\begin{equation}
    \begin{aligned}
        \phi_{TPMS} \geq c &\iff -\phi_{TPMS} \leq -c. \\
        -c \leq \phi_{TPMS} \leq c &\iff |\phi_{TPMS}| \leq c.
    \end{aligned}
\end{equation}

\begin{figure} [h] 
	\centering
	\includegraphics[width=0.98 \textwidth]{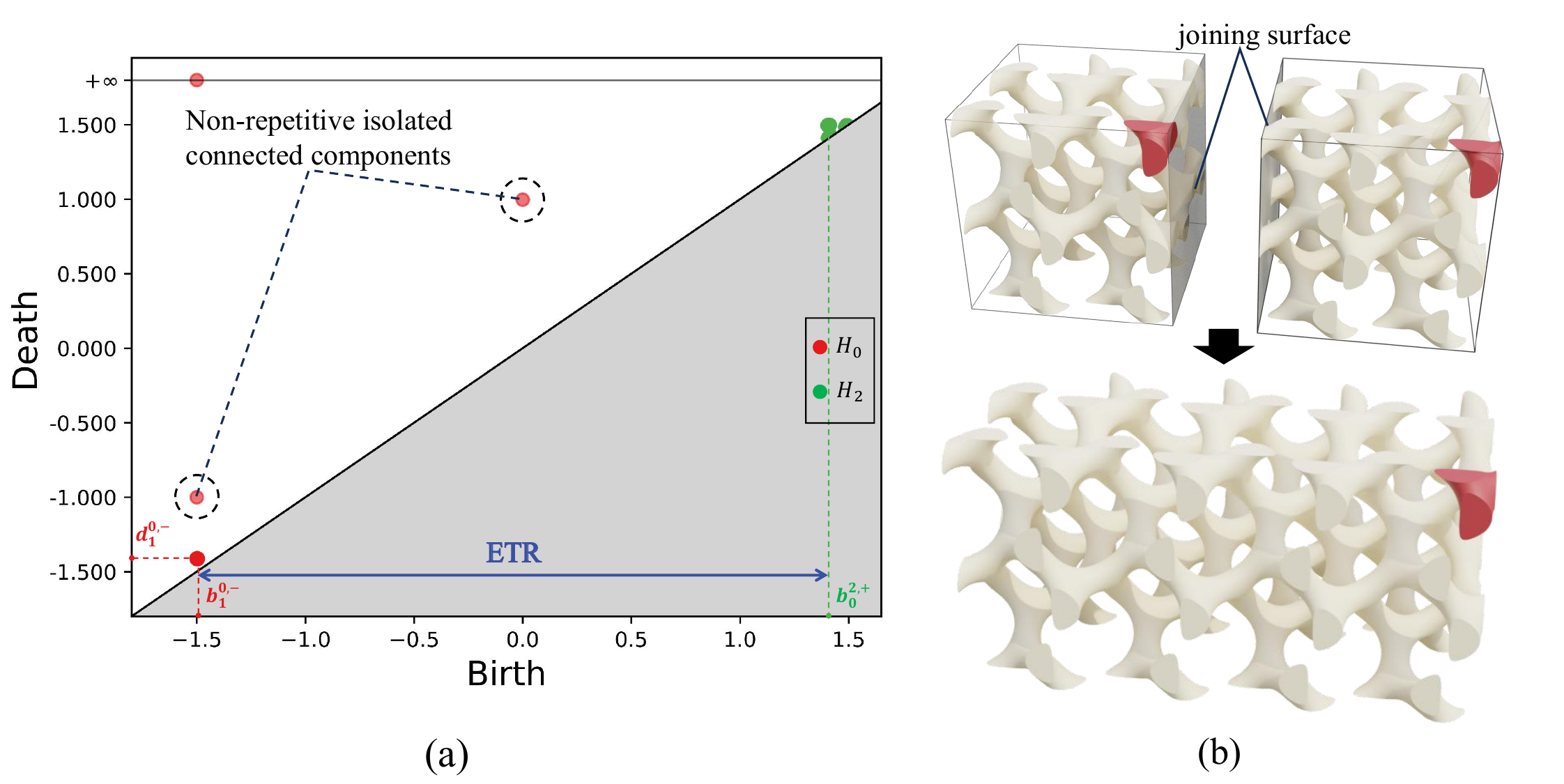}
	\caption{(a) Persistence diagram of Rod G-TPMS. (b) There is always only one non-repetitive isolated connected component in a porous structure jointed by units. }
	\label{fig: G_PD} 
\end{figure}

It is important to note that there is a specific situation for the G-TPMS structure. 
    Due to the periodic nature of TPMSs, if there is an isolated connected component in a complete unit, there will typically be the same isolated connected components in the other units.
    These connected components correspond to repetitive persistent pairs in the persistence diagram. 
    However, the isolated connected component (see Figure~\ref{fig: G_PD}(b)) in a complete unit of G-TPMS will be connected to solid materials of another unit during splicing.
    Therefore, the spliced G-type porous structure will have only one connected component (see Figure~\ref{fig: G_PD}(b)) and one non-overlapping persistent pair in the corresponding persistence diagram (see Figure~\ref{fig: G_PD}(a)).
    The isolated connected component that exists within a single unit but disappears during the splicing of units is referred to as a \textbf{non-repetitive isolated connected component}. 
    Additionally, isolated holes only form within the units or at the splicing boundary, and there are no instances of eliminating isolated holes during splicing.  
    Consequently, there are no occurrences of "non-repetitive isolated holes".

Although the non-repetitive isolated connected components cannot be manufactured due to their small size in the spliced porous structure, they should not be considered in the calculation and expansion of ETRs and EDRs. 
    Therefore, the ordering rules of the persistent pairs are changed. 
    Note that in a TPMS formed by at least two unit cells joined together in each direction, a non-repetitive isolated connected component corresponds to a non-overlapping persistent pair in the persistence diagram. 
    Conversely, other types of isolated connected components correspond to an overlapping persistent pair. 
    To ignore the effect of non-repetitive isolated connected components, if more than $\varsigma$ persistent pairs are present within a neighborhood of radius $\varepsilon$, then it participates in the sorting of the persistent pairs. 
    The $\varepsilon$ is set to $0.1$ and the $\varsigma$ is set to $1$ in the subsequent discussions. 
    As explained in Figure~\ref{fig: G_PD}(a), due to the absence of any other 0-dimensional persistent pairs within the neighborhood of two persistence pairs, they are not involved in the ordering of persistence pairs.

In conclusion, the calculation of ETRs of TPMSs can be realized using persistent homology. 
    Given the ETRs $[c_{min},c_{max}]$ of TPMSs, the porous structures correspond to the sets $\{\phi_{TPMS}\leq c_{min}\}$ and $\{\phi_{TPMS}\leq c_{max}\}$ can be converted into tetrahedral meshes using the marching tetrahedron method~\cite{doi1991efficient}. 
    Subsequently, the relative density $\rho_{min}$ and $\rho_{max}$ of these two porous structures can be approximated by dividing the volume of the corresponding tetrahedral mesh by the volume of the design domain. 
    The density range $[\rho_{min},\rho_{max}]$ is the resulting EDRs of TPMSs.

\subsection{Implicit B-spline representation of TPMSs}
\label{subsec: Parametric representation of TPMSs}
There is a corresponding relationship between the EDRs and ETRs of TPMSs. 
    Therefore, the EDRs can be extended by extending the ETRs.  
    As shown in Table~\ref{tab: TPMS}, expansion of the ETRs can only be achieved by adjusting the periodicity of TPMSs in three directions. 
    However, the limited adjustable parameters (periodicity in three directions) hinder the achievement of the desired result. 
    
B-splines are effective tools for representing complex geometric objects. 
    Gao et al.~\cite{gao2022connectivity} demonstrated the ability of B-spline functions to represent complex porous structures by utilizing them to interpolate a discrete distance field of a binary porous structure. 
    The trivariate B-spline function is employed as a new representation of TPMS by approximating a complete unit of the original TPMS. 
    Subsequently, the control coefficients of this B-spline function can be optimized to extend the ETR. 

Given a TPMS structure $\phi_{TPMS}=c$ with a periodicity of $2\pi$ in each direction, an $S\times S\times S$ number of data points are sampled from the function $\phi_{TPMS}$. 
    Specifically, positions $\{(x_i,y_i,z_i)\}_{i=0}^{S^3-1}$ are selected with equal intervals in the three directions within the range $[0,2\pi]\times [0,2\pi] \times [0,2\pi]$.
    These positions are then substituted into the function $\phi_{TPMS}$ to calculate the values $\{\xi_i\}_{i=0}^{S^3-1}$ for fitting. 
    The set of parameters corresponding to these data points is $\{(\frac{x_i}{2\pi},\frac{y_i}{2\pi},\frac{z_i}{2\pi})\}_{i=0}^{S^3-1}$. 
    Finally, the fitting problem is formulated as follows:
\begin{align}
    \label{eq: fitting problem}
        \min_{C_{ijk}}\quad &\sum_{i=0}^{S^3-1}| C(\frac{x_i}{2\pi},\frac{y_i}{2\pi},\frac{z_i}{2\pi}) - \xi_i) |^2  \\
        &C(u,v,w) = \sum_{i=0}^{n_u-1}\sum_{j=0}^{n_v-1}\sum_{k=0}^{n_w-1}R_{ijk}(u,v,w)C_{ijk}. \nonumber
\end{align}

In three-dimensional scenarios, where the number of data points is substantial, the least squares progressive-iterative approximation (LSPIA) algorithm~\cite{deng2014progressive} is used to solve the fitting problem due to its fast convergence speed and minimal memory usage. 
 
\subsection{Preservation of symmetry}
\label{subsec: Preservation of symmetry}
Solid TPMSs are cubic symmetric systems, which maintain the properties of TPMSs under transformations of coordinates such as reflections and rotations.
    This characteristic significantly reduces the computational burden of TPMSs in topology optimization~\cite{li2019design}. 
    However, fitting errors disrupt the symmetries of TPMSs, although the B-spline function used in Subsection~\ref{subsec: Parametric representation of TPMSs} approximates TPMSs well. 
    On the other hand, accurately approximating TPMS units requires numerous data points, leading to increased computational burden.
    In this section, a partial fitting method is introduced to address these two issues.

    By considering the rotation and reflection invariance of TPMSs, a complete unit of TPMS can be represented using a smaller unit. 
    For example, a P-TPMS with a period of $2\pi$ in all three directions is defined as:
    \begin{equation}
        \phi_{P}(x,y,z) = \cos x+\cos y+\cos z = c.
    \end{equation}
    Let $\phi_P|_{[0,2\pi]}$ be the function that restricts function $\phi_P$ to the domain $[0,2\pi]^3$. 
    The following holds for all $x\in [\pi,2\pi]$, and $y,z\in [0,\pi]$:
    \begin{equation}
        \begin{aligned}
            \label{eq: transfer of P-TPMS}
            \phi_P|_{[0,2\pi]}(x,y,z) =& \cos x+\cos y+cos z \\
            =& \cos(2\pi-x)+\cos y+cos z\\
             =& \phi_P|_{[0,\pi]}(2\pi-x,y,z).
        \end{aligned}
    \end{equation}
Equation~\ref{eq: transfer of P-TPMS} applies equally to situations where $y\in [\pi,2\pi]$ and $z\in [\pi,2\pi]$. 
    A reflection function $\eta^X$ is hereby defined:
    \begin{equation}
        \eta^X(x) \triangleq  
        \begin{cases}
            x,\quad 0\leq x \leq X \\
            2X - x,\quad X < x \leq 2X.
        \end{cases}
    \end{equation}
    To simplify, $(\eta^X(x),\eta^X(y),\eta^X(z))$ is denoted as $\eta^X(x,y,z)$ in the following discussions.
    A complete P-TPMS unit can be generated using a half unit as follows:
\begin{equation}
    \begin{aligned}
        \phi_P|_{[0,2\pi]}(x,y,z) = \phi_P|_{[0,\pi]}(\eta^\pi(x,y,z)).
    \end{aligned}        
\end{equation}

The periodicity of TPMSs allows for representing the structure of the entire space with one complete unit. 
    A translation function $\iota^X$ is defined as follows:
\begin{equation}
    \begin{aligned}
        \iota^X :~\mathbb{R} &\rightarrow [0,2X) \\
        x &\mapsto x-2X\lfloor\frac{x}{2X}\rfloor.
    \end{aligned}
\end{equation}
To simplify, $(\iota^X(x),\iota^X(y),\iota^X(z))$ is denoted as $\iota^X(x,y,z)$ in the following discussions. 
    Subsequently, a P-TPMS structure with any size can be generated using a half unit (representing the structure within $[0,\pi]^3$):
\begin{equation}
    \begin{aligned}
        \label{eq: infinite P-TPMS}
         \phi_{P}(x,y,z) 
        =& \cos x+\cos y+cos z \\
        =& \cos(\iota^\pi(x)) + \cos(\iota^\pi(y)) + \cos(\iota^\pi(z)) \\
        =& \phi_P|_{[0,2\pi]}(\iota^\pi(x,y,z)) \\
        =& \phi_P|_{[0,\pi]}(\eta^\pi\circ\iota^\pi(x,y,z)).
    \end{aligned}
\end{equation}

Instead of fitting the data points sampled from $[0,2\pi]^3$ as illustrated in Subsection~\ref{subsec: Parametric representation of TPMSs}, the data points sampled from $[0,\pi]^3$ can be fitted to obtain the implicit B-spline representation $C_P(u,v,w)$ of the P-TPMS. 
    Subsequently, the P-TPMS under implicit B-spline representation within the whole space is defined by reflecting and translating.
    It is important to note that the parametric domain of $C_P(u,v,w)$ is $[0,1]^3$, corresponding to the P-TPMS located in $[0,\pi]^3$.
    Therefore, the period in each direction of the P-TPMS under implicit B-spline representation is $2$. 
    A P-TPMS function $F_P$ defined in $\mathbb{R}^3$ can be constructed using the same approach as Equation~\ref{eq: infinite P-TPMS}:
    \begin{equation}
        \begin{aligned}
        \label{eq: parametric TPMS}
            F_{P}(x,y,z) 
            =& F_P|_{[0,2]}(\iota^1(x,y,z)) \\
            =& F_P|_{[0,1]}(\eta^1\circ\iota^1(x,y,z))\\
=&C_P(\eta^1\circ\iota^1(x,y,z)).
        \end{aligned}
    \end{equation} 
    Equation~\ref{eq: parametric TPMS} shows that a TPMS represented by a B-spline function can be obtained by fitting a half unit of this TPMS. 
    The computational burden is reduced because fitting a smaller structure requires fewer data points. 
    Additionally, the function $\eta^X$ denotes reflection. 
    As a result, the new P-TPMS fucntion $F_P$, as defined by Equation~\ref{eq: parametric TPMS}, is unchanged by reflection. 
    The errors induced in approximations do not affect the symmetry of the P-TPMS under implicit B-spline representation. 
 %-------------------------------------------------------------------------
\subsection{Generation of extended TPMSs}
\label{subsec: Method for expanding the effective threshold range}
After obtaining TPMSs represented by B-spline functions, the objective of this subsection is to optimize the control coefficients of the B-splines to extend the ETRs and EDRs while preserving the original TPMS structure as much as possible.

To achieve this, a topological objective function that aims to optimize the ETRs and further extend the EDRs is established.  
    It should be noted that when the porous structure, which is spliced by $2\times 2\times 2$ complete units, does not have any additional connected components or isolated holes, due to periodicity, the structure spliced by this unit in the entire space will also not have any additional connected components or isolated holes.
    Therefore, the topological features of $2\times 2\times 2$ units are calculated, which corresponds to the original P-TPMS function $\phi_P|_{[0,4\pi]}$ and the new P-TPMS function $F_P|_{[0,4]}$. 
    If there are no isolated holes and additional connected components within the $[0,4]^3$ domain, then the porous structure in the entire space will also not have any additional connected components or isolated holes. 
    Consequently, in the following discussion, $F_P|_{[0,4]}$ is used as a real-valued function to induce filtration for computing the topological features of TPMSs. 
    
One intuitive method to extend the ETRs is by setting a loss function as follows:
\begin{equation}
    \label{eq: bad loss function}
    L_0  =-(b_0^{2,+} - d_1^{0,-}).
\end{equation}
    Although this loss function can extend the ETR in the experiments, it is difficult to achieve convergence.
    This is because there are no restrictions on the expansion of the ETR, which results in solutions that deviate significantly from the original TPMS structure. 
    Therefore, an upper bound on the expansion is needed to be imposed.
\begin{comment}
Supposing that a parametric TPMS meets $\hat{F}_P(x,y,z) \leq -M,~\forall (x,y,z)\in \mathbb{R}^3$, where $M$ is a large positive number. 
    Then, the valid threshold range of this parameteric TPMS is $[-M,+\infty)$. 
    However, the structure $\{\hat{F}_P(x,y,z)\leq -M\}$ is a solid cube, obviously, the function $\hat{F}_P(x,y,z)$ is very different with the initial TPMS.
    The loss function in Equation~\ref{eq: bad loss function} will lead such a result.
\end{comment} 
     
Assuming that the ETR of the original TPMS is $[c^0_{min},c^0_{max}]$, the length of the ETR is $l_0 =c^0_{max}- c^0_{min}$.
    The ETR is aimed to be extended to $[c^0_{min}-\mu l_0,c^0_{max}+\mu l_0] \triangleq [c_{min},c_{max}]$, where $\mu \in \mathbb{R}^+$ is the expansion ratio. 
    Then, the loss function can be defined as follows: 
    \begin{equation}
        \begin{aligned}
            L_{top} = (d_1^{0,-} - c_{min})^2 + (~b_0^{2,+}- c_{max})^2.
        \end{aligned}
    \end{equation}

In addition to extending the ETR, another goal is to make the extended TPMS as similar as possible to the original TPMS within the original ETR $\{c_{min}^0\leq \phi_P \leq c_{max}^0\}$.
   Let the set $\{c_{min}^0\leq \phi_P \leq c_{max}^0\}$ be denoted by $\mathcal{A}$. 
   The loss function to maintain the similarity is defined as follows:
\begin{equation}
    \begin{aligned}
        L_{sim} = 
    \int_{\mathcal{D}}
    \left(
        C_P(u,v,w)-\phi_P(\pi u,\pi v,\pi w)
    \right)^2 \cdot \mathcal{X}_{\mathcal{A}}(\pi u,\pi v,\pi w)d\omega,
    \end{aligned}
\end{equation}
where $\mathcal{D}$ is the parametric domain $[0,1]^3$, and $\mathcal{X}_{\mathcal{A}}$ is the indicator function defined by:
\begin{equation}
    \mathcal{X}_{\mathcal{A}}(u,v,w) = 
    \begin{cases}
        0,\quad (u,v,w)\notin \mathcal{A}\\
        1,\quad (u,v,w)\in \mathcal{A}.
    \end{cases}
\end{equation}

Due to the inability to obtain an accurate expression of $\mathcal{X}_{\mathcal{A}}$, a B-spline function is employed to approximate it. 
    $I\times I\times I$ points are sampled with equal intervals within $\mathcal{D}$ to generate a set $\{(u_i,v_i,w_i)\}_{i=0}^{I^3-1}$. 
    Then, the corresponding values form a set $\{\mathcal{X}_{\mathcal{A}}(\pi u_i,\pi v_i,\pi w_i)\}_{i=0}^{I^3-1}$. 
    These data points are also fitted using the LSPIA algorithm to obtain a B-spline function $g_{\mathcal{A}}(u,v,w)$. 
    The loss function can be approximated as follows:
    \begin{equation}
        \begin{aligned}
            \tilde{L}_{sim} = 
            \int_{\mathcal{D}}
            \left(
                C_P(u,v,w)-\phi_P(\pi u,\pi v,\pi w)
            \right)^2 \cdot g_{\mathcal{A}}(u,v,w)d\omega.
        \end{aligned}
    \end{equation}

The final loss function is defined as:
\begin{equation}
    L = (1-\alpha) L_{top} + \alpha \tilde{L}_{sim},
\end{equation} 
where $\alpha$ is the weight. 

Using the topological inverse mapping $\tilde{\pi}_{F_P}$ introduced in Equation~\ref{eq: map persistent pair to 0-simplexes}, the 0-simplices can be obtained as follows:
\begin{equation}
    \begin{aligned}
        (\sigma_1^{0},\tau_1^{0}) &= \tilde{\pi}_{F_P}(b_1^{0,-},d_1^{0,-}) \\
        (\sigma_0^{2},\tau_0^{2}) &= \tilde{\pi}_{F_P}(b_0^{2,+},d_0^{2,+}).
    \end{aligned}
\end{equation}
It should be noted that $\tau_1^{0}$ and $\sigma_0^{2}$ are 0-simplices, corresponding to vertices in $\mathbb{R}^3$, so they can be directly substituted into the function for calculation.
    Based on the topological inverse mapping, the derivative of the loss function $L_{top}$ is derived as follows:
\begin{equation}
    \begin{aligned}
        \label{eq: Ltop}
        \frac{\partial L_{top}}{\partial C_{ijk}}  &= 2(d_1^{0,-} - c_{min})\frac{\partial d_1^{0,-}}{\partial C_{ijk}} + 2(~b_0^{2,+}- c_{max})\frac{\partial b_0^{2,+}}{\partial C_{ijk}}\\
    &= 2(d_1^{0,-} - c_{min})\frac{\partial F_P(\tau_1^{0})}{\partial C_{ijk}} + 2(~b_0^{2,+}- c_{max})\frac{\partial F_P(\sigma_0^{2})}{\partial C_{ijk}} \\
    &= 2(d_1^{0,-} - c_{min})\frac{\partial C_P(\eta^1\circ\iota^1(\tau_1^{0}))}{\partial C_{ijk}} + 2(~b_0^{2,+}- c_{max})\frac{\partial C_P(\eta^1\circ\iota^1(\sigma_0^{2}))}{\partial C_{ijk}}\\
    &= 2(d_1^{0,-} - c_{min})R_{ijk}(\eta^1\circ\iota^1(\tau_1^{0}))+ 2(~b_0^{2,+}- c_{max})R_{ijk} (\eta^1\circ\iota^1(\sigma_0^{2}))
    \end{aligned}
\end{equation}

The derivative of the loss function $\tilde{L}_{sim}$ can be calculated as follows:
\begin{equation}
    \begin{aligned}
        \label{eq: Lsim}
    \frac{\partial\tilde{L}_{sim}}{\partial C_{ijk}} = 2\int_{\mathcal{D}}
    &\left(
        C_P(u,v,w)-\phi_P(\pi u,\pi v,\pi w)
    \right) \\ &\cdot g_{\mathcal{A}}(u,v,w)\cdot R_{ijk}(u,v,w)d\omega,
    \end{aligned}
\end{equation}
where $R_{ijk}(u,v,w)$ is the blending basis function. 

Finally, the optimization of the loss function $L$ can be carried out using the adaptive gradient descent method~\cite{duchi2011adaptive} based on Equation~\ref{eq: Ltop} and Equation~\ref{eq: Lsim}. 
    The optimized function $F_P$ represents an implicit porous structure, which is referred to as an extended TPMS.
%-------------------------------------------------------------------------

\section{Implementation and experiments}
\label{sec: Implementation and discussion}
The proposed design method is implemented using the C++ programming language and tested on a PC with a 2.90 GHz i7-10700 CPU and 16 GB RAM. 
    This section presents the effectiveness of the proposed method and provides experimental details.

\subsection{Errors under the parametric representation} 
\label{subsec: Errors under the parametric representation}

In Section~\ref{subsec: Parametric representation of TPMSs}, the TPMSs are converted from a periodic nodal representation to an implicit B-spline representation using a partial fitting.
    Consequently, it is necessary to discuss the errors between these two representations. 

\begin{figure} [h] 
	\centering
	\subfloat[]{
		\includegraphics[width=0.24\textwidth]{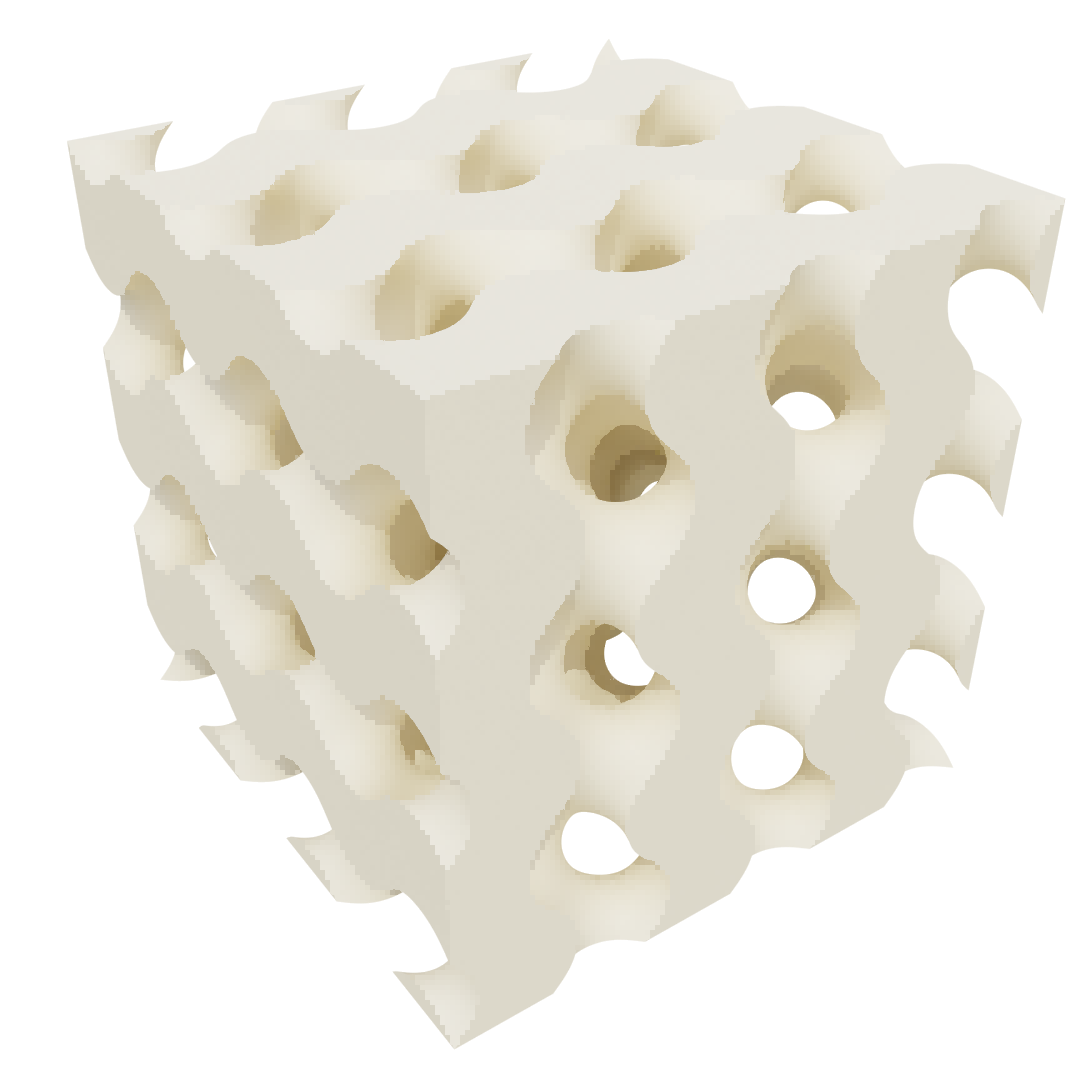}} 
    \subfloat[]{
		\includegraphics[width=0.24\textwidth]{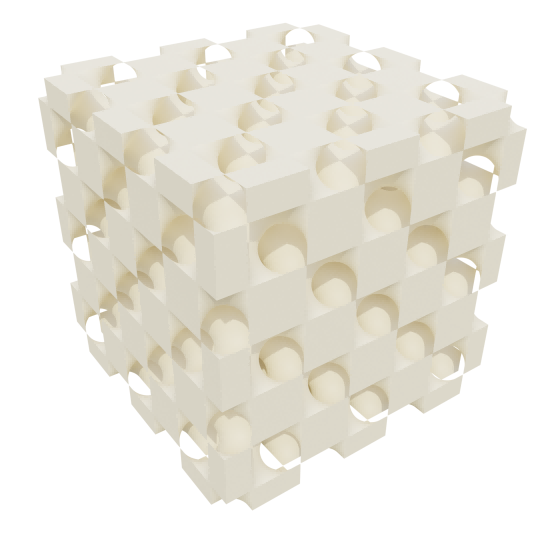}}
    \subfloat[]{
		\includegraphics[width=0.24\textwidth]{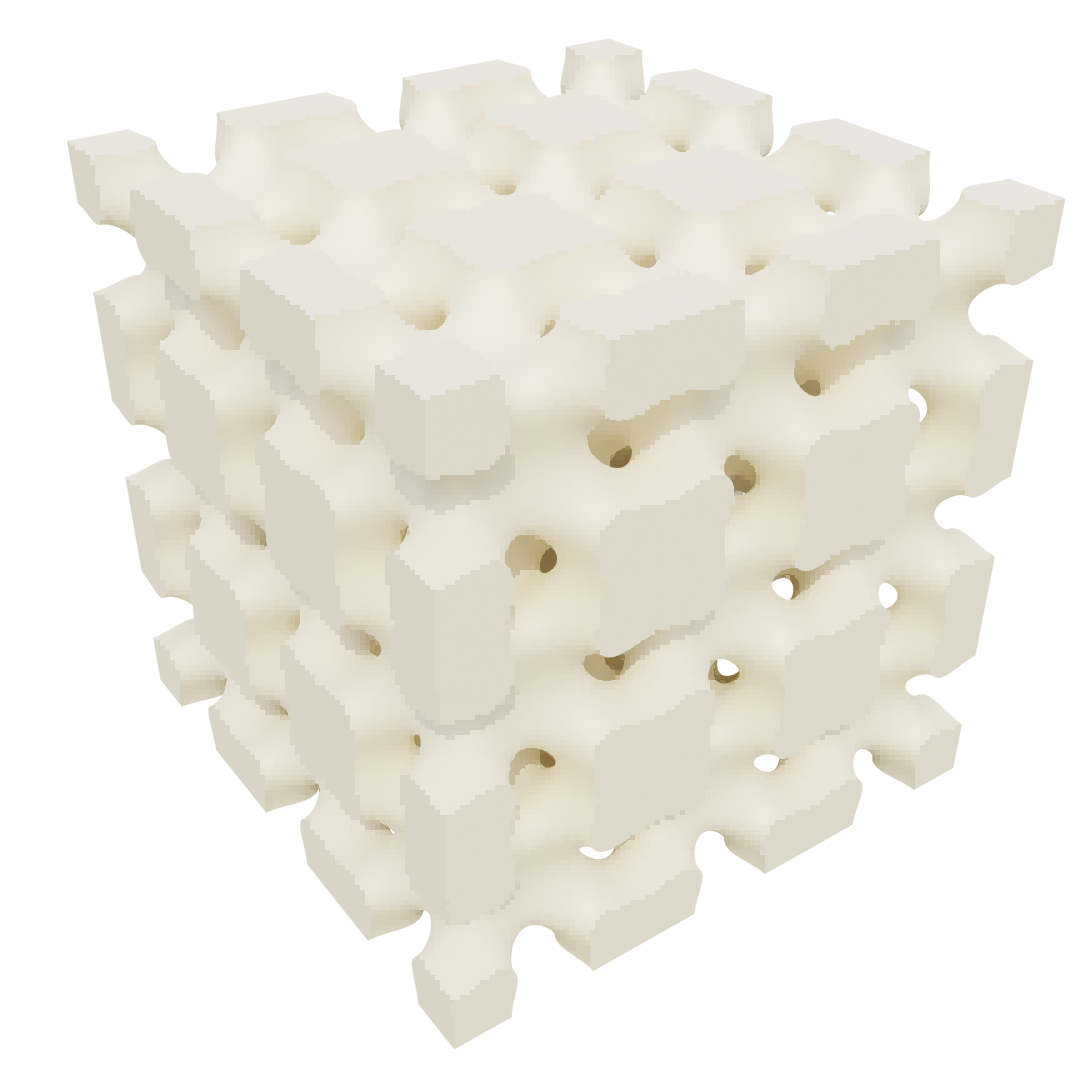}} \\
    \subfloat[]{
		\includegraphics[width=0.24\textwidth]{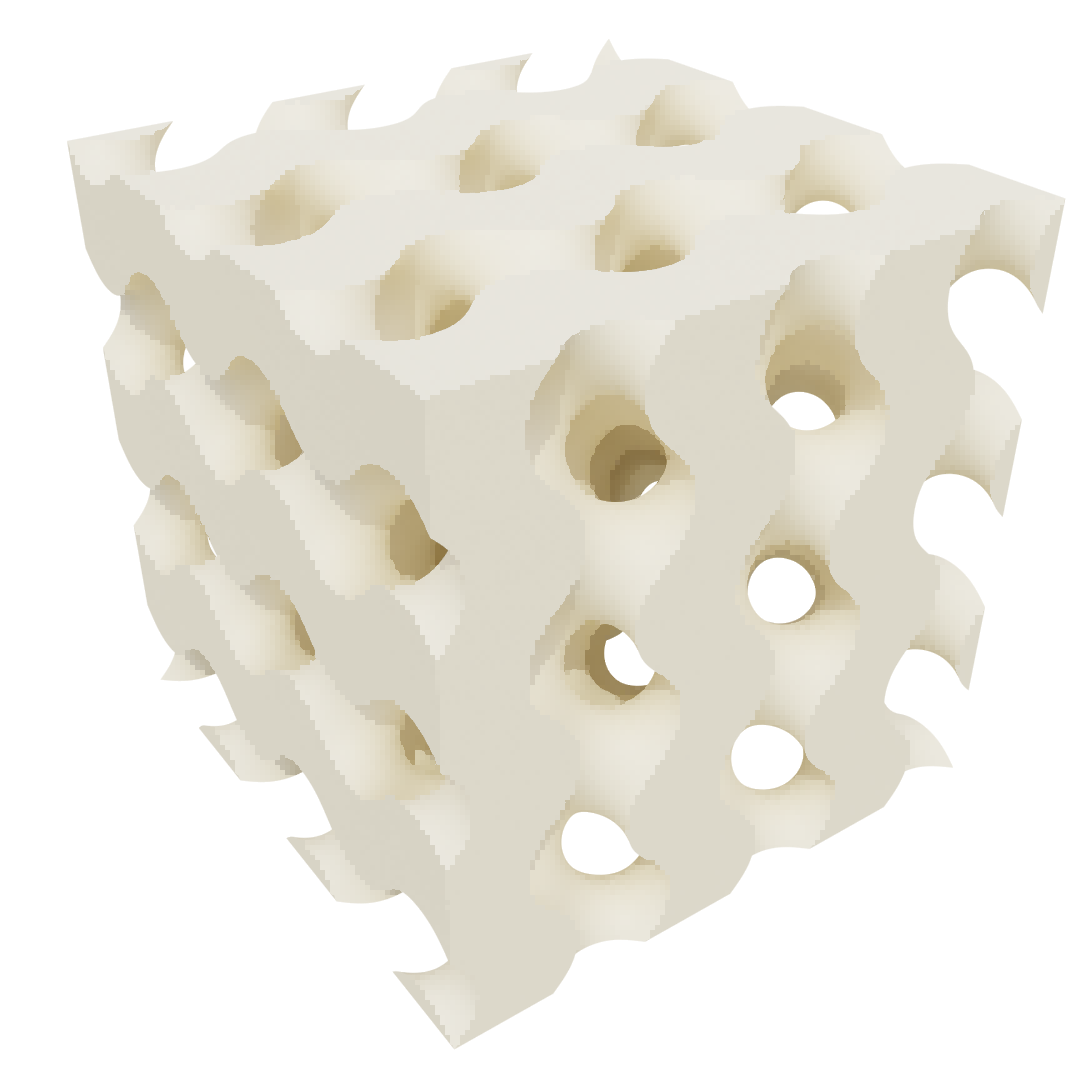}} 
    \subfloat[]{
		\includegraphics[width=0.24\textwidth]{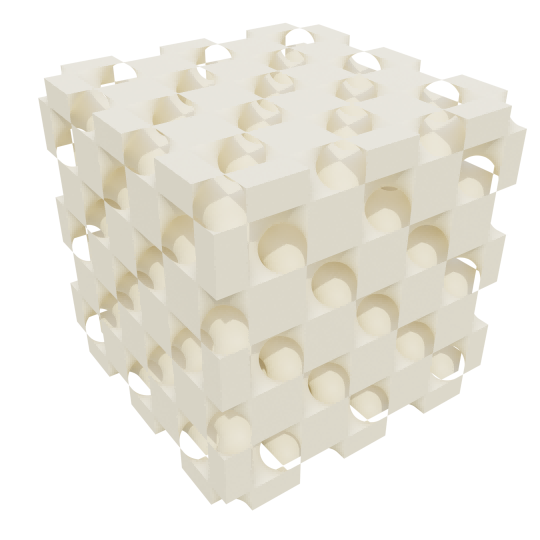}}
    \subfloat[]{
		\includegraphics[width=0.24\textwidth]{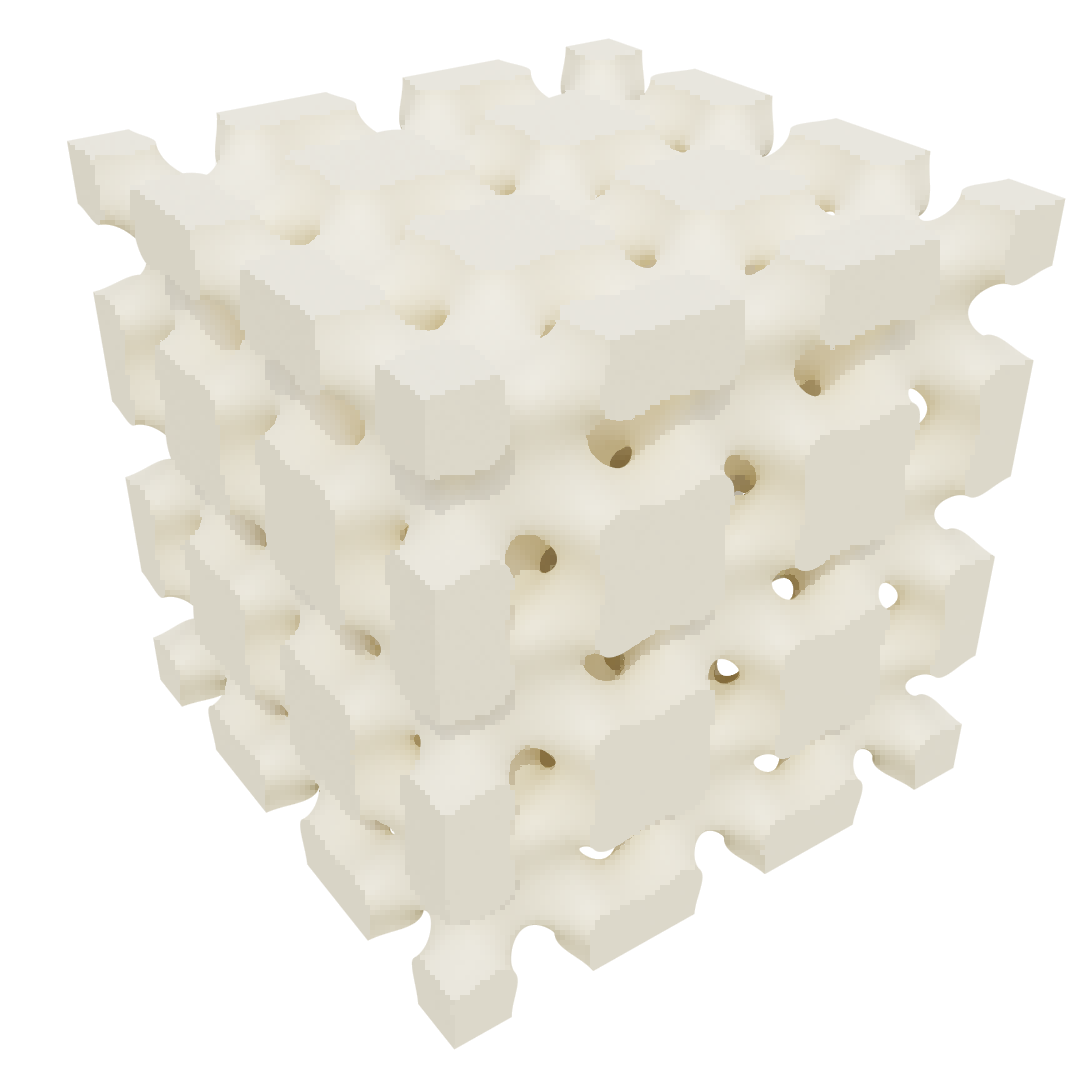}}
	\caption{Comparison of Rod TPMSs under implicit B-spline representation and periodic nodal representation. The first row shows TPMSs under periodic nodal representation, while the second row shows TPMSs represented by B-spline functions. (a) and (d) demonstrate G-TPMSs. (b) and (e) demonstrate D-TPMSs. (c) and (f) demonstrate I-WP-TPMSs. }
	\label{fig: Comparison of TPMS solid} 
\end{figure}

From a visualization perspective, as illustrated in Figure~\ref{fig: Comparison of TPMS solid}, there is no significant difference between solid structure $\{\phi_{TPMS}\leq c\}$ and $\{F_{TPMS}\leq c\}$, where $\phi_{TPMS}$ and $F_{TPMS}$ represent the periodic nodal representation (see Table~\ref{tab: TPMS}) and implicit B-spline representation (see Equation~\ref{eq: parametric TPMS}), respectively.

In Section~\ref{subsec: Preservation of symmetry}, the half unit of TPMS is fitted and then the symmetry and periodicity of P-TPMS are utilized to obtain a P-TPMS under implicit B-spline representation that extends throughout the entire space. 
    This approach reduces fitting errors compared to fitting the complete TPMS unit and then using periodicity to define a function across the entire space.  
    These two fitting methods are referred to as the \textbf{partial fitting method} and the \textbf{complete fitting method}, respectively. 
    The Mean Square Error (MSE) for both fitting methods under varying numbers of control coefficients is presented in Table~\ref{table: MSE of fitting}.
    The MSE decreases as the number of control coefficients increases. 
    Furthermore, the MSE of the partial fitting method is lower than the complete fitting method when the same number of control coefficients is used.  
    Although increasing the number of control coefficients can reduce the MSE, it also increases the computational complexity.      
    Therefore, considering that the fitting error is sufficiently small in the partial fitting method, $10\times 10\times 10$ control coefficients are used in the following experiments to balance both factors.

\begin{table}[]
\centering
\caption{The mean squared errors (1e-6) of fitting periodic nodal TPMSs.}
\label{table: MSE of fitting}
\resizebox{0.98\textwidth}{!}{
\begin{tabular}{cccclccc}
\hline
         & \multicolumn{3}{c}{Complete fitting method}                                 &  & \multicolumn{3}{c}{Partial fitting method}                                     \\ \cline{2-4} \cline{6-8} 
         & $8\times 8\times 8$ & $10\times 10\times 10$ & $12\times 12\times 12$ &  & $8\times 8\times 8$ & $10\times 10\times 10$ & $12\times 12\times 12$ \\ \cline{1-4} \cline{6-8} 
P-type   & 1906.10              & 433.44                  & 105.85                  &  & 129.48               & 18.78                   & 5.29                   \\
D-type   & 9684.09             & 301.17                  & 137.30                 &  & 263.60               & 36.75                   & 11.47                   \\
G-type   & 619.15             & 253.41                  & 45.24                   &  & 22.99                & 9.93                   & 4.81                  \\
IWP-type & 2304.19            & 792.97                 & 84.56                  &  & 8.31                & 2.47                   & 0.26                   \\ \hline
\end{tabular}}
\end{table}

In conclusion, the error between the implicit B-spline representations and periodic nodal representations of TPMSs is extremely small.  
    Additionally, the proposed symmetry-preserving method can effectively reduce the error.

\subsection{Optimization results}
\label{subsec: Optimization results}  
In this subsection, the effectiveness of the proposed method is demonstrated by comparing the extended TPMSs with the original ones.

\begin{figure}[h]
    \centering
    \includegraphics[width=.98\textwidth]{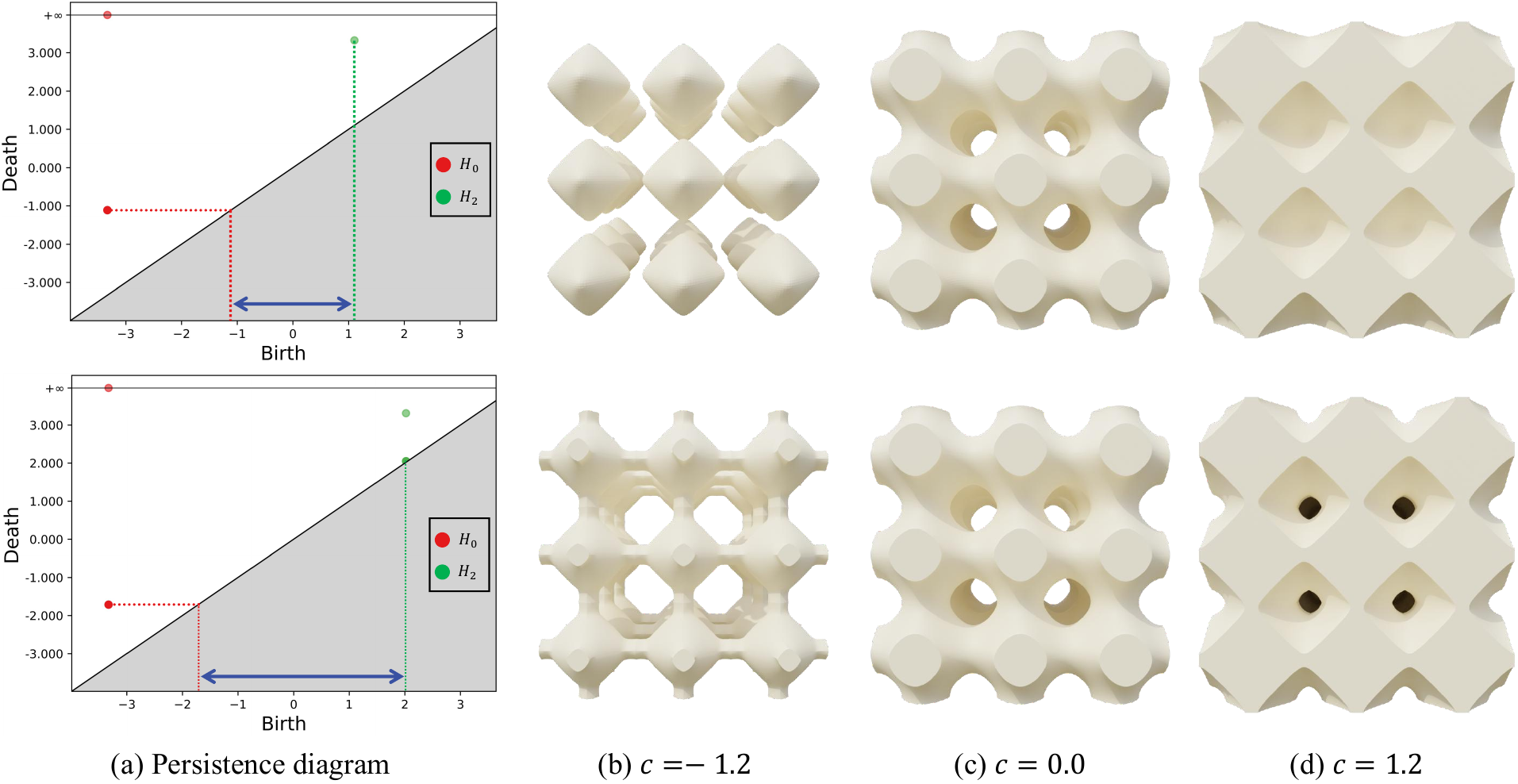}
        \caption{Optimization of Rod P-TPMS. The first row is the solid P-TPMS $\{\phi_P \leq c\}$ under periodic nodal representation. The second row is the extended P-TPMS represented by a trivariate B-spline function after optimization. The blue arrows in the PDs represent the ETRs.}
    \label{fig: P_optimization}
\end{figure}

The ETR for the Rod P-TPMS, represented by $\Phi_P\leq c$ under periodic nodal representation, is $[-1.113,~ 1.105]$. 
    The corresponding EDR is $[0.207,~0.776]$. 
    When $c < -1.113$, the P-TPMS exhibits multiple connected components as depicted in Figure~\ref{fig: P_optimization}(b).  
    Similarly, isolated holes appear in the solid structure when $c \geq 1.105$ (see Figure~\ref{fig: P_optimization}(d)).
    After transforming to implicit B-spline representation and optimizing, the ETR extends to $[-1.714,2.034]$. 
    Furthermore, the EDR is extended to $[0.110,0.916]$. 
    As shown in the second row of Figure~\ref{fig: P_optimization}, the additional connected components and isolated holes are successfully eliminated after optimization.  
    Simultaneously, as illustrated in Figure~\ref{fig: P_optimization}(b), when $c$ falls within the initial ETR, there is no significant difference in the extended P-TPMS and original P-TPMS.  

\begin{figure}[h]
    \centering
    \includegraphics[width=.98\textwidth]{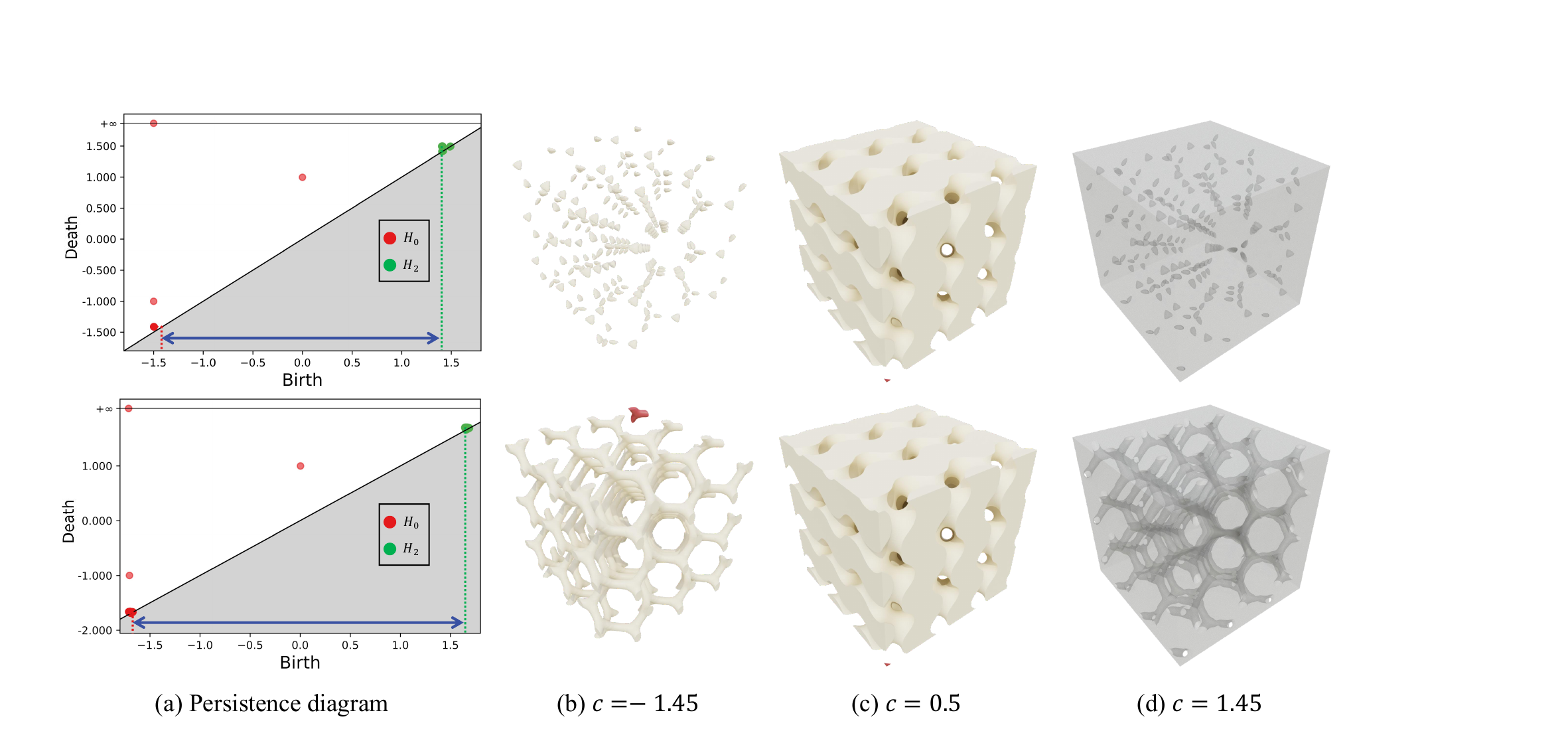}
        \caption{Optimization of Rod G-TPMS. The fourth column is depicted in a perspective view for easy observation of hole distribution. The first row is the Rod G-TPMS $\{\phi_P \leq c\}$ under periodic nodal representation. The second row is the resulting extended G-TPMS represented by a function $\{F_P \leq c\}$ after optimization. The blue arrows in the PDs represent the ETRs. The non-repetitive isolated connected components are marked in red.}
    \label{fig: G_optimization}
\end{figure}

The ETR and EDR of the Rod G-TPMS are $[-1.41,1.40]$ and $[0.018,0.979]$, respectively. 
    Its ETR is extended with a ratio of $0.1$ 
to show the effectiveness of the proposed method. 
    The optimized results are illustrated in Figure~\ref{fig: G_optimization}. 
    The ETR and the corresponding EDR of the G-TPMS are extended to $[-1.66,1.67]$ and $[0.008,0.993]$, respectively. 
    As shown in Figure~\ref{fig: G_optimization}(d), the numerous isolated holes in the G-TPMS are eliminated in the extended G-TPMS. 
    Additionally, when $c$ falls within the initial ETR, the extended G-TPMS exhibits no distinct difference from the original G-TPMS (see Figure~\ref{fig: G_optimization}(c)).
    The numerous isolated connected components in the first row of Figure~\ref{fig: G_optimization}(b) are merged into a principal connected component and a small non-repetitive isolated connected component.

Since the non-repetitive isolated connected components are not considered in the order of persistent pairs, the resulting PD still has two corresponding persistent pairs. 
    Moreover, as illustrated in Figure~\ref{fig: G_optimization}(b) and Figure~\ref{fig: G_optimization}(c), the non-repetitive isolated connected components still exist in the extended G-TPMS, marked in red. 
    Due to their tiny size compared to the porous structure, they have no significant effect. 
    On the other hand, optimizing the non-repetitive isolated connected components drastically changes the structure of the initial G-TPMS, resulting in prominent differences between the extended G-TPMS and the original G-TPMS. 
    Therefore, considering the reasons mentioned above, it is better to neglect the non-repetitive isolated connected components during ordering.

\subsection{Effect of parameters}
\label{subsec: Effect of parameters}
In this subsection, the influence of parameters on the optimization results is analyzed. 
    Specifically, the learning rate $\eta$ of the adaptive gradient descent, the weight $\alpha$, and the expansion ratio $\mu$ of the loss function are analyzed. 

The aim of the optimization is to maintain the structure $\mathcal{A} = \{c_{min}^0\leq \phi_P \leq c_{max}^0\}$, which corresponds to the original ETR $[c_{min}^0,c_{max}^0]$ and to change structures outside this range. 
    To quantify the variation of set $\mathcal{A}$ after optimization, $N$ points are sampled from $\mathcal{A}$, resulting in a set of coordinates $\{(u_k,v_k,w_k)\}_{k=0}^{N-1}$.
    Subsequently, the quantified function $E_{sim}$ is defined as:
\begin{equation}
    \begin{aligned}
        E_{sim} = \frac{1}{N-1}\sum_{k=0}^{N-1}[F_P(u_k,v_k,w_k)-\phi_P(\pi u_k,\pi v_k,\pi w_k)]^2.
    \end{aligned}
\end{equation}
    Typically, points are uniformly sampled from the parametric domain $[0,1]^3$ to obtain the set of coordinates $\{(u_k,v_k,w_k)\}_{k=0}^{N-1}$ and any points that are not in $\mathcal{A}$ are filtered out.

\begin{figure}[h]
    \centering
    \includegraphics[width=0.8\textwidth]{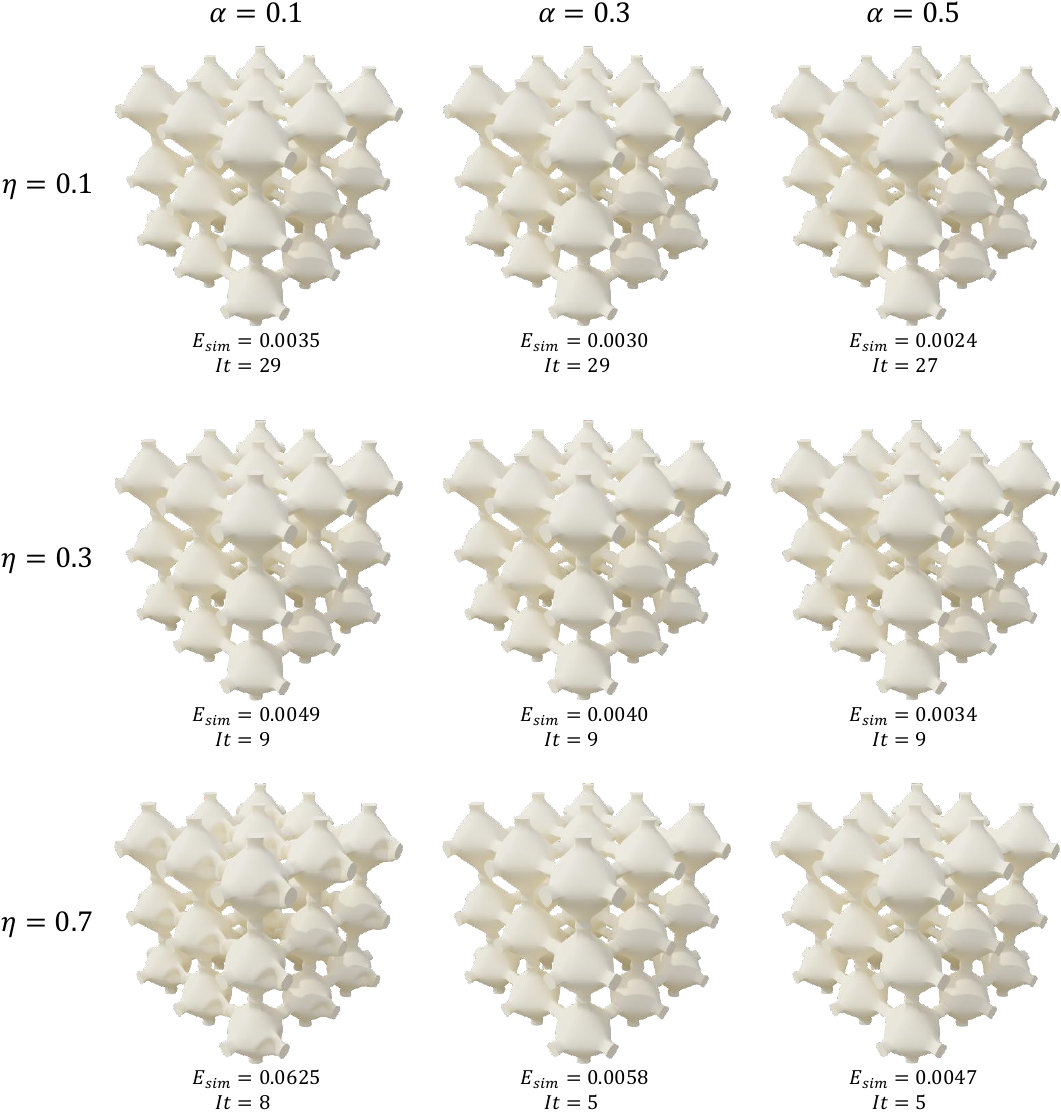}
    \caption{A comparison of experimental results with varying learning rates $\eta$ and weights $\alpha$ is conducted to analyze their impact. The difference between the TPMSs before and after optimization is quantified by $E_{sim}$, while $It$ represents the number of iterations. }
    \label{fig: parameter effect}
\end{figure}

Initially, the expansion ratio is fixed, while the learning rate and weight are varied. 
   Figure~\ref{fig: parameter effect} illustrates that the number of iterations required for convergence decreases with an increasing learning rate $\eta$, while the error $E_c$ decreases with an increasing weight $\alpha$. 
   However, as the learning rate increases, the error $E_c$ also increases. 
   Meanwhile, as shown in the first column of the third row in Figure~\ref{fig: parameter effect}, when the learning rate is very high, strange shapes are formed in the optimized structure. 
   As shown in the second and third columns of the third row in Figure~\ref{fig: parameter effect}, increasing the weight can suppress the generation of such strange structures. 
   To achieve a balance between similarity and iteration numbers, unless otherwise specified, the learning rate is set to 0.3 and the weight is set to 0.5 in the following experiments.

\begin{figure} [h] 
	\centering
	\subfloat[$\mu=0.1$]{
		\includegraphics[width=0.3\textwidth]{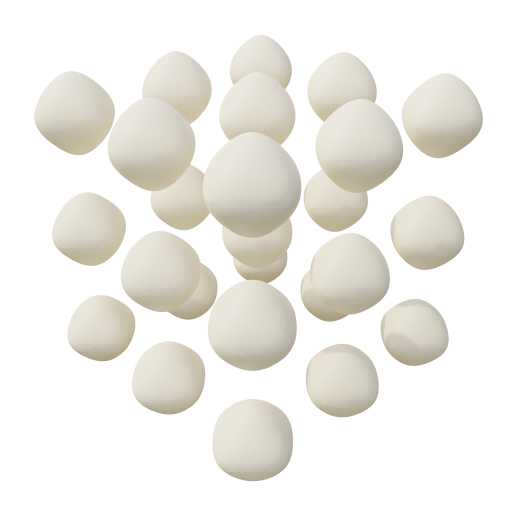}} 
	\subfloat[$\mu=0.3$]{
		\includegraphics[width=0.3\textwidth]{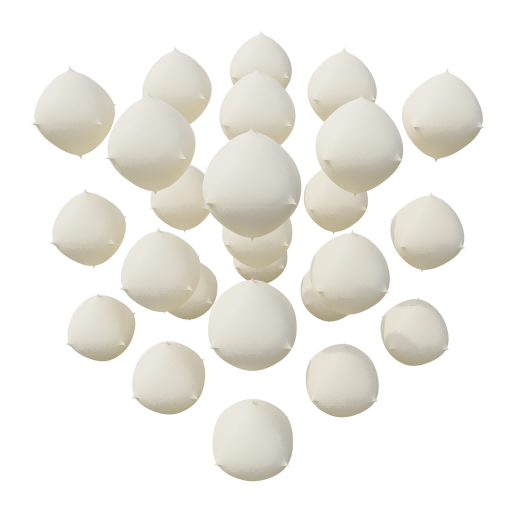}}
	\subfloat[$\mu=0.5$]{ 
		\includegraphics[width=0.3\textwidth]{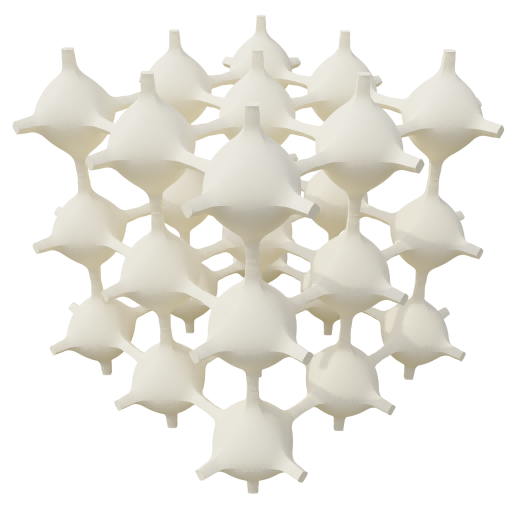}}
	\caption{Extended P-TPMSs $\{F_p \leq 1.5\}$ under different values of expansion ratio $\mu$.}
	\label{fig: effect of expansion ratio} 
\end{figure}

Subsequently, the effect of the expansion ratio $\mu$ on the resulting structures is discussed.
    Given an original Rod P-TPMS with an ETR of $[-1.113,1.105]$, the ETR of the extended P-TPMS is shown in Table~\ref{table: effect of expansion ratio}. 
    As the expansion ratio $\mu$ increases, the ETR of the extended P-TPMS also extends. 
    However, the similarity quantified by the error $E_{sim}$ also increases simultaneously.  
    Figure~\ref{fig: effect of expansion ratio} visualizes the resulting structures $\{F_P \leq -1.5\}$.  
    With the growth of the expansion ratio, the isolated connected components gradually merge into a single connected component.

\begin{table}[h]
\centering
\caption{Effect of expansion ratio $\mu$. The effective threshold range (ETR), effective relative density range (EDR), and similarity error ($E_{sim}$) of the extended P-TPMSs under different expansion ratio $\mu$ are recorded.}
\resizebox{0.58\textwidth}{!}{
\begin{tabular}{cccc}
\hline
                          & $\mu = 0.1$      & $\mu = 0.3$      & $\mu = 0.5$      \\ \hline
ETR & $[-1.334,1.325]$ & $[-1.637,1.754]$ & $[-1.816,2.109]$ \\
EDR   & $[0.165,0.821]$  & $[0.118,0.885]$  & $[0.102,0.925]$  \\
$E_{sim}$                 & 0.0004           & 0.0034           & 0.0085           \\ \hline
\end{tabular}}
\label{table: effect of expansion ratio}
\end{table}

\subsection{Analysis of symmetry}
\label{subsec: Analysis of symmetry}
The periodic nodal TPMS with a periodicity of $2\pi$ in the domain $[0, 2\pi]^3$ can be represented by a TPMS in the domain $[0, \pi]^3$.
    Using this property, the complete unit can be reconstructed by employing a B-spline function obtained from fitting the half unit of the TPMS.  
    Experiments in Subsection~\ref{subsec: Errors under the parametric representation} show that this fitting method significantly reduces the fitting error. 
    This subsection aims to demonstrate that this fitting strategy maintains the cubic symmetry of TPMSs during optimization and fitting. 

Given an FRD-TPMS with a period of $2\pi$, B-spline functions $\tilde{C}_{FRD}$ and $C_{FRD}$ are employed to fit the complete unit and half unit of the FRD-TPMS, respectively. 
    Then, function $\tilde{F}_{FRD}$ is defined by $\tilde{C}_{FRD}$ based on the periodicity. 
    Meanwhile, function $F_{FRD}$ is defined by $C_{FRD}$ based on the symmetry and periodicity.
    These two functions are then optimized to extend the EDRs using the method described in Subsection~\ref{subsec: Method for expanding the effective threshold range}. 
    The solid structures corresponding to $\tilde{F}_{FRD}$ and $F_{FRD}$ are referred to as the \textbf{asymmetric structure} and the \textbf{symmetric structure}, respectively. 
    
\begin{figure*}[t]
    \centering
    \includegraphics[width=.98\textwidth]{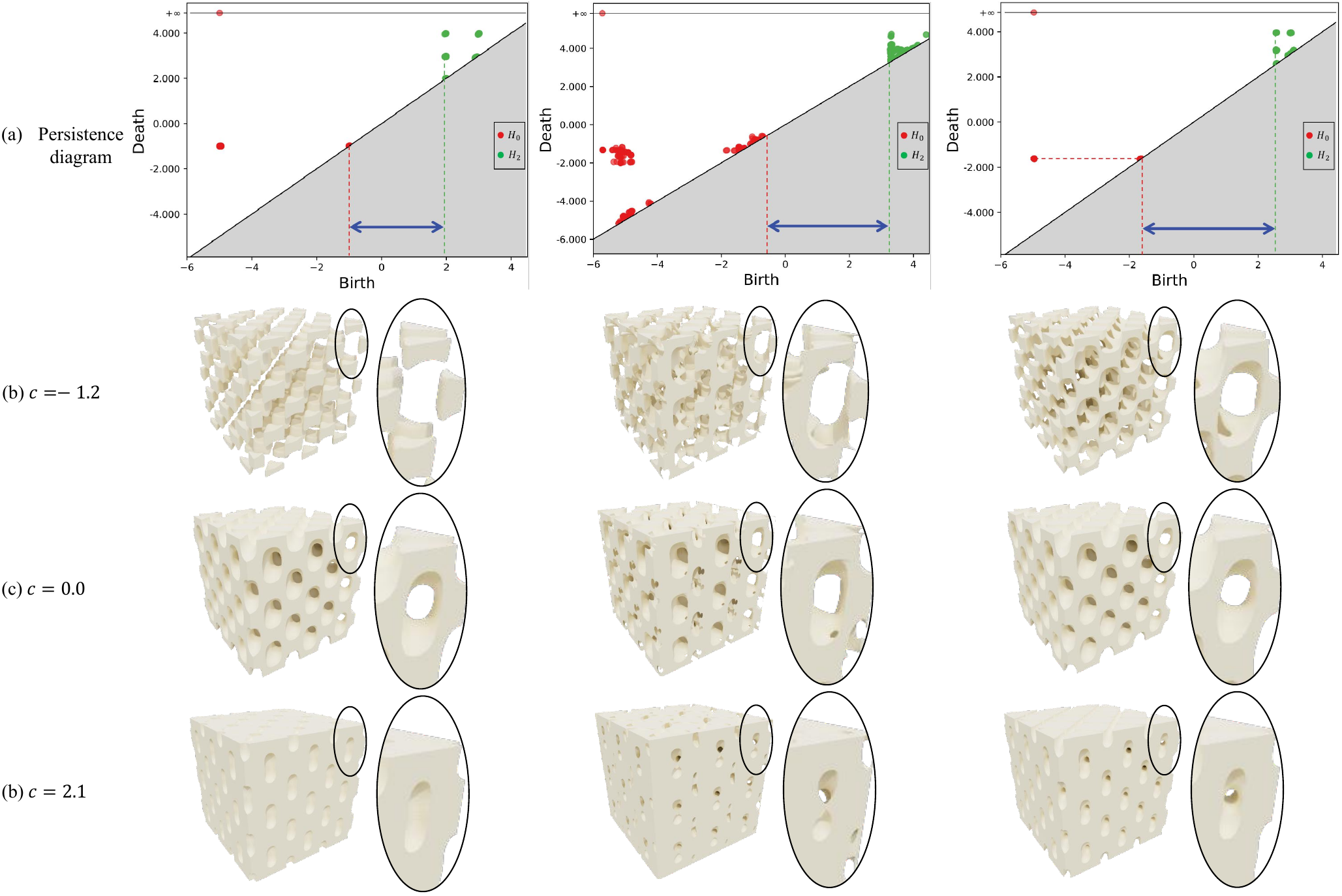}
    \caption{Optimization of FRD-TPMS under implicit B-spline representation. The first column represents the solid FRD-TPMS $\{\phi_{FRD} \leq c\}$ under periodic nodal representation. The second column is the resulting asymmetric FRD-TPMS represented by a function $\{\tilde{F}_{FRD} \leq c\}$. The third column is the resulting symmetric FRD-TPMS represented by a function $\{F_{FRD} \leq c\}$. The blue arrows in the PDs represent the ETR.}
    \label{fig: FRD_optimization}
\end{figure*}

The first column row of Figure~\ref{fig: FRD_optimization} depicts the original FRD-TPMS represented by a periodic nodal formula.
    The second column of Figure~\ref{fig: FRD_optimization} demonstrates that the implicit B-spline representation obtained by fitting the complete unit does not retain the symmetry of the original FRD-TPMS after optimization. 
    In contrast, the third column of Figure~\ref{fig: FRD_optimization} illustrates that the implicit B-spline representation obtained by the partial fitting method maintains the cubic symmetry of the original FRD-TPMS after optimization.  
    The B-spline function $C_{TPMS}$ is obtained by fitting half unit. 
    Subsequently, the porous structure is generated using symmetry and periodicity. 
    Although the B-spline function $C_{TPMS}$ is adjusted during optimization, the symmetry and periodicity are maintained, which is impossible with the complete fitting method. 
    Therefore, the PD of the optimized symmetric structure contains a few overlapping persistent pairs. 
    On the other hand, the PD of the asymmetric structure exhibits overlapping persistent pairs due to its periodicity, but many persistent pairs are gathered without overlap due to its asymmetry. 
    In conclusion, these results demonstrate that the method successfully preserves the symmetry of the TPMS during optimization and fitting, while extending its ETR and EDR.

\subsection{Comparison with periodic nodal representation of TPMSs}
\label{subsec: Comparison with periodic nodal representation of TPMSs}
A TPMS with a wider EDR has a larger design domain.
    This subsection presents a demonstration of the advantages of the extended P-TPMS over the original P-TPMS in topology optimization through numerical simulations and mechanical tests.

Given force and displacement boundary conditions, the topology optimization method aims to find the best distribution of the material under a volume constraint to minimize the compliance of the model. 
    The mathematical formula of this problem is as follows:
\begin{equation}
    \begin{aligned}
    \label{eq:topology optimization}
    \min_{\rho(u,v,w)} \quad & \mathcal{C}(\rho)=U^T K U \\
    \text{s.t.} \quad & KU=F  \\
                      & K= \sum_{e=1}^{N_e}\int_{\Omega_e}B_e^T C^H_{TPMS}(\rho) B_e d\Omega  \\
                      & V = \int_{\Omega}\rho(u,v,w)d\Omega = \vartheta  \text{Vol}(\Omega)  \\
                      & \varrho_{min}\leq \rho(u,v,w) \leq \varrho_{max} ,   
    \end{aligned}
\end{equation}
where the compliance $\mathcal{C}(\rho)$ is the objective function, the density distribution $\rho(u,v,w)$ is the optimization variable, $K$ is the global stiffness matrix, $U$ is the displacement, $F$ is the load vector, $C^H_{TPMS}$ is the homogenized elastic tensor of TPMSs which is calculated using the homogenization method~\cite{andreassen2014determine}, $N_e$ is the number of elements, $V$ is the volume of the porous model, $\vartheta$ is the fraction of the volume, Vol$(\Omega)$ is the volume of the design domain, and $[\varrho_{min},\varrho_{max}]$ is the constraint range of density.     

The EDR of a Rod P-TPMS under periodic nodal representation is $[0.21, 0.78]$. 
    To avoid the occurrence of overly fine structures constraints are added in Equation~\ref{eq:topology optimization} as $\varrho_{min} = 0.21+0.05=0.26$ and $\varrho_{max} = 0.78$. 
    The EDR of the extended P-TPMS is $[ 0.11 , 0.91 ]$. 
    For the topology optimization of the extended P-TPMS, $\varrho_{min}$ and $\varrho_{max}$ are defined as $\varrho_{min} = 0.11 + 0.05 = 0.16$ and $\varrho_{max} = 0.91$. 
    Since the solution of equation~\ref{eq:topology optimization} does not need to consider the manufacturing of the structure, the density constraint of P-TPMS under periodic nodal representation is also adjusted to $[0.16,0.91]$ for the purpose of comparison in topology optimization. 
    The optimization problem defined by Equation~\ref{eq:topology optimization} is solved using optimality criteria algorithm~\cite{bendsoe2003topology}. 
    
\begin{figure}[h]
    \centering
    \includegraphics[width=.98\textwidth]{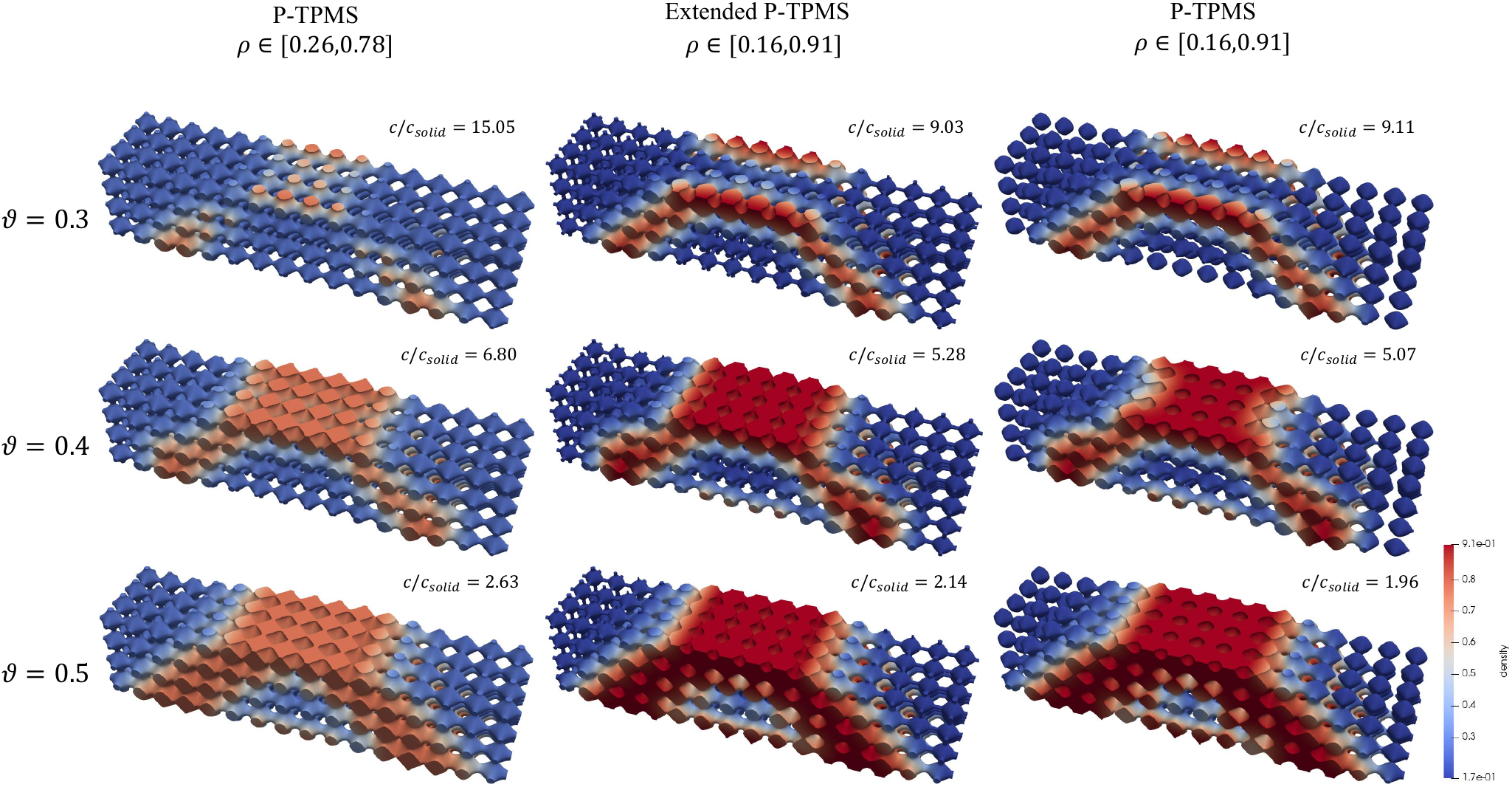}
        \caption{Porous models obtained through topology optimization. The red color represents high relative density, while the blue color represents low relative density. Each column presents optimized models with different volume ratios under the same density range and unit type. Each row shows optimization results with different relative density ranges and unit types under the same volume ratio. The normalized compliance is listed at the top right of each model.}
    \label{fig: TO_result}
\end{figure}

\begin{figure}[h]
    \centering
    \includegraphics[width=.98\textwidth]{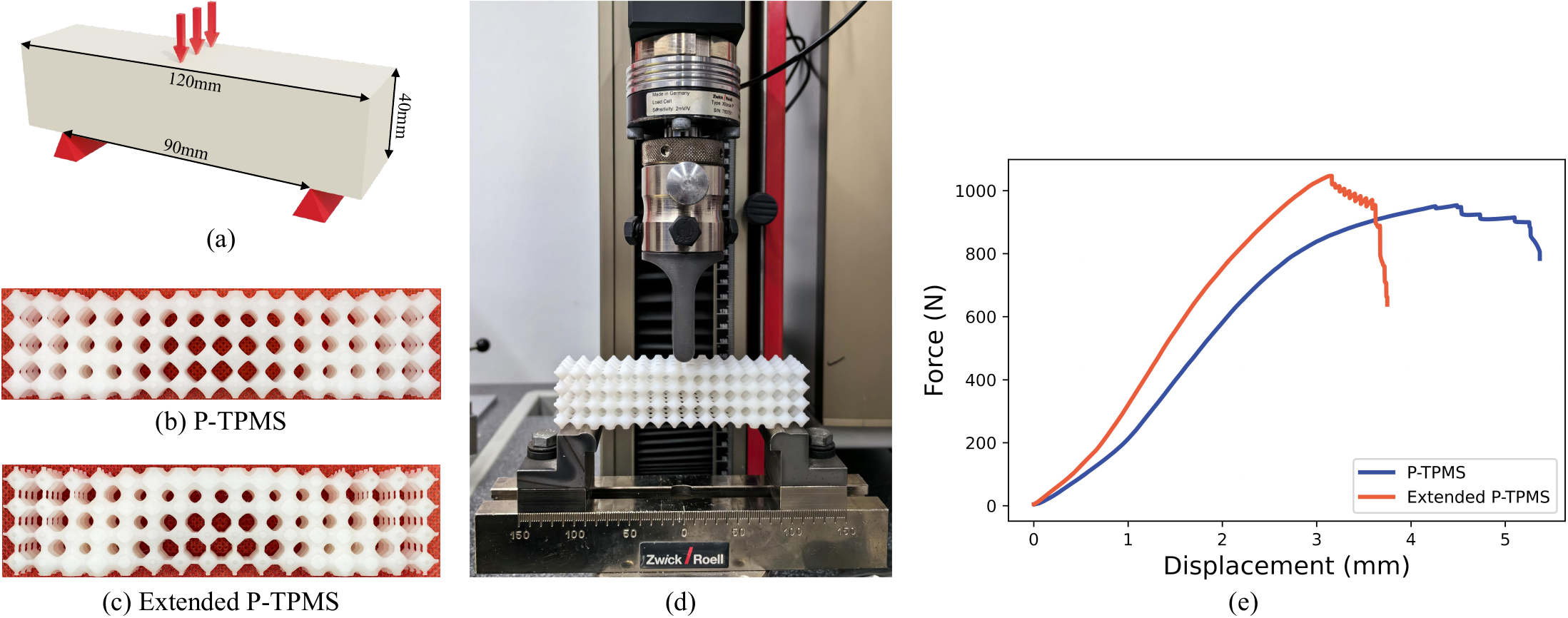}
        \caption{(a) Experimental model (b) The manufactured P-TPMS model. (c) The manufactured extended P-TPMS model. (d) Experimental setup. (e) Force-displacement curve.}
    \label{fig: TO_setup}
\end{figure}

The boundary conditions of the experiment are depicted in Figure~\ref{fig: TO_setup}(a). 
    The beam has dimensions of $120\times120\times120$mm.
    A uniform force of $500$N is applied from the top.
    Figure~\ref{fig: TO_result} displays the optimized models and the normalized compliance (porous model compliance divided by solid model compliance) under the same boundary conditions.
    Comparing the first and third columns in Figure~\ref{fig: TO_result} under the same volume constraint, it is evident that a wider range of relative density constraints leads to better solutions for the topology optimization problem.  
    Since the range of relative density constraints set in the third column exceeds the EDR of the P-TPMS, a large number of isolated connected components appear in the low-density region, making the models unmanufacturable. 
    The extended P-TPMS not only improves the EDR but also maintains similarity to the original P-TPMS. 
    Therefore, the extended P-TPMS achieves minimum compliance while ensuring manufacturability. 
    In the first row, the EDR lower bound of the P-TPMS is very close to the volume constraint of 0.3, which means that most materials cannot reach the upper bound of the density constraint (0.78) in order to satisfy the volume constraint. 
    However, the extended P-TPMS, with its higher EDR, can place more high-relative-density materials in areas that require reinforcement. 
    Therefore, as the volume constraint decreases, the improvement of the extended P-TPMS compared to the P-TPMS in the topology optimization problem becomes more significant.

To further demonstrate the correctness of the proposed method, a porous model with a volume ratio of 0.3 is created using additive manufacturing, as shown in Figure~\ref{fig: TO_setup}. 
    The dimensions of the model are $120\times 120\times 120$mm.
    It is worth noting that the extended P-TPMS closely resembles the original P-TPMS, indicating a high degree of similarity. 
    Since P-TPMS has self-supporting properties, there was no need for support material during the manufacturing process of all porous models. 
    Figure~\ref{fig: TO_setup} showcases the samples obtained from the experiment, while Figure~\ref{fig: TO_setup}(d) and Figure~\ref{fig: TO_setup}(e) depict the experimental setup and results, respectively. 
    The results in Figure~\ref{fig: TO_setup}(e) demonstrate that the extended P-TPMS exhibits higher stiffness compared to the original P-TPMS in the optimized models. 
    This outcome further reinforces the effectiveness of the proposed method.

\section{Conclusion and future work}
\label{sec: Conclusion}
Porous structures with a wider EDR offer a larger design domain for model designs. 
    To calculate and extend the EDRs, the EDRs of TPMSs are analyzed from a topological perspective using persistent homology. 
    By converting TPMSs into implicit B-spline representation through fitting, the EDRs can be extended by optimizing a topological objective function, resulting in extended TPMSs.
    These extended TPMSs are then used in topology optimization problems to design heterogeneous porous models with high stiffness. 
    Experimental results demonstrate that the extended TPMSs have higher stiffness compared to the original TPMSs due to their larger EDRs.
    Although the experiments only show certain types of extended TPMSs, the method is applicable to other TPMSs defined by the equation $\phi_{TPMS} = c$. 
    This study introduces a novel representation of TPMSs using B-spline functions, providing more adjustable parameters (control coefficients of B-spline functions) than the original TPMSs (periodicity and threshold). 
    Therefore, this makes it possible to optimize the EDRs of TPMSs with implicit B-spline representations, which is a challenging task for traditional TPMSs. 
    Given their adjustability, extended EDRs, and high stiffness, these extended TPMSs could become a better option for designing porous models.

This study only discusses the equivalence between optimizing Pore and Sheet type TPMSs, and Rod type TPMSs from a theoretical standpoint. 
    More experiments are needed to confirm this theory.
    Additionally, the relationship between complete and half units of TPMSs is deduced to keep symmetry and improve fitting accuracy. 
    Notably, complete TPMS units can be constructed from smaller units due to their inherent symmetry. 
    Future studies will aim to preserve this symmetry in extended TPMSs by defining a complete unit within a smaller-than-half unit. 
    Furthermore, with numerous adjustable parameters in extended TPMSs, future work includes developing an objective function and optimizing control coefficients to achieve designs such as self-supporting heterogeneous models and constant mean curvature models. 

\section*{CRediT authorship contribution statement}
\textbf{Depeng Gao:} Conceptualization, Methodology, Software, Validation, Writing - Original Draft, Writing - Review \& Editing, Visualization. \textbf{Yuanzhi Zhang:} Validation, Data Curation. \textbf{Hongwei Lin:} Writing - Review \& Editing, Supervision, Methodology, Conceptualization.    

\section*{Acknowledgments}
\noindent
This work is supported by National Natural Science Foundation of China under Grant nos. 62272406 and 61932018.

\section*{Declaration of competing interest}
The authors declare that they have no known competing financial interests or personal relationships that could have appeared to influence the work reported in this paper.

\section*{Declaration of Generative AI and AI-assisted technologies in the writing process}
During the preparation of this work the author(s) used ChatGPT in order to improve language and readability. After using this tool/service, the author(s) reviewed and edited the content as needed and take(s) full responsibility for the content of the publication.

\bibliographystyle{elsarticle-num} 
\bibliography{ref}

\begin{thebibliography}{10}\itemsep=-1pt

\bibitem{andreassen2014determine}
E.~Andreassen and C.~S. Andreasen.
\newblock How to determine composite material properties using numerical
  homogenization.
\newblock {\em Computational Materials Science}, 83:488--495, 2014.

\bibitem{bendsoe2003topology}
M.~P. Bendsoe and O.~Sigmund.
\newblock {\em Topology optimization: theory, methods, and applications}.
\newblock Springer Science \& Business Media, 2003.

\bibitem{bruel2020topology}
R.~Br{\"u}el-Gabrielsson, V.~Ganapathi-Subramanian, P.~Skraba, and L.~J.
  Guibas.
\newblock Topology-aware surface reconstruction for point clouds.
\newblock In {\em Computer Graphics Forum}, volume~39, pages 197--207. Wiley
  Online Library, 2020.

\bibitem{chazal2021introduction}
F.~Chazal and B.~Michel.
\newblock An introduction to topological data analysis: fundamental and
  practical aspects for data scientists.
\newblock {\em Frontiers in artificial intelligence}, 4:108, 2021.

\bibitem{chen2020porous}
H.~Chen, Q.~Han, C.~Wang, Y.~Liu, B.~Chen, and J.~Wang.
\newblock Porous scaffold design for additive manufacturing in orthopedics: a
  review.
\newblock {\em Frontiers in Bioengineering and Biotechnology}, 8:609, 2020.

\bibitem{deng2014progressive}
C.~Deng and H.~Lin.
\newblock Progressive and iterative approximation for least squares b-spline
  curve and surface fitting.
\newblock {\em Computer-Aided Design}, 47:32--44, 2014.

\bibitem{dong2022topology}
Z.~Dong, J.~Chen, and H.~Lin.
\newblock Topology-controllable implicit surface reconstruction based on
  persistent homology.
\newblock {\em Computer-Aided Design}, 150:103308, 2022.

\bibitem{duchi2011adaptive}
J.~Duchi, E.~Hazan, and Y.~Singer.
\newblock Adaptive subgradient methods for online learning and stochastic
  optimization.
\newblock {\em Journal of machine learning research}, 12(7), 2011.

\bibitem{edelsbrunner2002topological}
Edelsbrunner, Letscher, and Zomorodian.
\newblock Topological persistence and simplification.
\newblock {\em Discrete \& Computational Geometry}, 28:511--533, 2002.

\bibitem{feng2018review}
J.~Feng, J.~Fu, Z.~Lin, C.~Shang, and B.~Li.
\newblock A review of the design methods of complex topology structures for 3d
  printing.
\newblock {\em Visual Computing for Industry, Biomedicine, and Art},
  1(1):1--16, 2018.

\bibitem{feng2018porous}
J.~Feng, J.~Fu, C.~Shang, Z.~Lin, and B.~Li.
\newblock Porous scaffold design by solid t-splines and triply periodic minimal
  surfaces.
\newblock {\em Computer Methods in Applied Mechanics and Engineering},
  336:333--352, 2018.

\bibitem{feng2019sandwich}
J.~Feng, J.~Fu, C.~Shang, Z.~Lin, and B.~Li.
\newblock Sandwich panel design and performance optimization based on triply
  periodic minimal surfaces.
\newblock {\em Computer-Aided Design}, 115:307--322, 2019.

\bibitem{feng2022triply}
J.~Feng, J.~Fu, X.~Yao, and Y.~He.
\newblock Triply periodic minimal surface (tpms) porous structures: From
  multi-scale design, precise additive manufacturing to multidisciplinary
  applications.
\newblock {\em International Journal of Extreme Manufacturing}, 4(2):022001,
  2022.

\bibitem{feng2021isotropic}
J.~Feng, B.~Liu, Z.~Lin, and J.~Fu.
\newblock Isotropic porous structure design methods based on triply periodic
  minimal surfaces.
\newblock {\em Materials \& Design}, 210:110050, 2021.

\bibitem{feng2022stiffness}
Y.~Feng, T.~Huang, Y.~Gong, and P.~Jia.
\newblock Stiffness optimization design for tpms architected cellular
  materials.
\newblock {\em Materials \& Design}, 222:111078, 2022.

\bibitem{gao2022connectivity}
D.~Gao, J.~Chen, Z.~Dong, and H.~Lin.
\newblock Connectivity-guaranteed porous synthesis in free form model by
  persistent homology.
\newblock {\em Computers \& Graphics}, 106:33--44, 2022.

\bibitem{gao2023free}
D.~Gao, H.~Lin, and Z.~Li.
\newblock Free-form multi-level porous model design based on truncated
  hierarchical b-spline functions.
\newblock {\em Computer-Aided Design}, 162:103549, 2023.

\bibitem{hu2023isogeometric}
C.~Hu, H.~Hu, H.~Lin, and J.~Yan.
\newblock Isogeometric analysis-based topological optimization for
  heterogeneous parametric porous structures.
\newblock {\em Journal of Systems Science and Complexity}, 36(1):29--52, 2023.

\bibitem{hu2021heterogeneous}
C.~Hu and H.~Lin.
\newblock Heterogeneous porous scaffold generation using trivariate b-spline
  solids and triply periodic minimal surfaces.
\newblock {\em Graphical Models}, 115:101105, 2021.

\bibitem{kaczynski2004computational}
T.~Kaczynski, K.~M. Mischaikow, and M.~Mrozek.
\newblock {\em Computational homology}, volume~3.
\newblock Springer, 2004.

\bibitem{li2019design}
D.~Li, N.~Dai, Y.~Tang, G.~Dong, and Y.~F. Zhao.
\newblock Design and optimization of graded cellular structures with triply
  periodic level surface-based topological shapes.
\newblock {\em Journal of Mechanical Design}, 141(7):071402, 2019.

\bibitem{li2018optimal}
D.~Li, W.~Liao, N.~Dai, G.~Dong, Y.~Tang, and Y.~M. Xie.
\newblock Optimal design and modeling of gyroid-based functionally graded
  cellular structures for additive manufacturing.
\newblock {\em Computer-Aided Design}, 104:87--99, 2018.

\bibitem{li2021simple}
Y.~Li, Q.~Xia, S.~Yoon, C.~Lee, B.~Lu, and J.~Kim.
\newblock Simple and efficient volume merging method for triply periodic
  minimal structures.
\newblock {\em Computer Physics Communications}, 264:107956, 2021.

\bibitem{liu2023multiscale}
H.~Liu, L.~Chen, Y.~Jiang, D.~Zhu, Y.~Zhou, and X.~Wang.
\newblock Multiscale optimization of additively manufactured graded
  non-stochastic and stochastic lattice structures.
\newblock {\em Composite Structures}, 305:116546, 2023.

\bibitem{montemurro2022thermal}
M.~Montemurro, K.~Refai, and A.~Catapano.
\newblock Thermal design of graded architected cellular materials through a
  cad-compatible topology optimisation method.
\newblock {\em Composite Structures}, 280:114862, 2022.

\bibitem{ozdemir2023novel}
M.~Ozdemir, U.~Simsek, G.~Kiziltas, C.~E. Gayir, A.~Celik, and P.~Sendur.
\newblock A novel design framework for generating functionally graded
  multi-morphology lattices via hybrid optimization and blending methods.
\newblock {\em Additive Manufacturing}, 70:103560, 2023.

\bibitem{piegl1996nurbs}
L.~Piegl and W.~Tiller.
\newblock {\em The NURBS book}.
\newblock Springer Science \& Business Media, 1996.

\bibitem{pinho2009asymptotic}
J.~Pinho-da Cruz, J.~Oliveira, and F.~Teixeira-Dias.
\newblock Asymptotic homogenisation in linear elasticity. part i: Mathematical
  formulation and finite element modelling.
\newblock {\em Computational Materials Science}, 45(4):1073--1080, 2009.

\bibitem{poulenard2018topological}
A.~Poulenard, P.~Skraba, and M.~Ovsjanikov.
\newblock Topological function optimization for continuous shape matching.
\newblock In {\em Computer Graphics Forum}, volume~37, pages 13--25. Wiley
  Online Library, 2018.

\bibitem{shi2021design}
X.~Shi, W.~Liao, T.~Liu, C.~Zhang, D.~Li, W.~Jiang, C.~Wang, and F.~Ren.
\newblock Design optimization of multimorphology surface-based lattice
  structures with density gradients.
\newblock {\em The International Journal of Advanced Manufacturing Technology},
  117:2013--2028, 2021.

\bibitem{xu2023topology}
W.~Xu, P.~Zhang, M.~Yu, L.~Yang, W.~Wang, and L.~Liu.
\newblock Topology optimization via spatially-varying tpms.
\newblock {\em IEEE Transactions on Visualization and Computer Graphics}, 2023.

\bibitem{yang2014multi}
N.~Yang, Z.~Quan, D.~Zhang, and Y.~Tian.
\newblock Multi-morphology transition hybridization cad design of minimal
  surface porous structures for use in tissue engineering.
\newblock {\em Computer-Aided Design}, 56:11--21, 2014.

\bibitem{yoo2012heterogeneous}
D.-J. Yoo.
\newblock Heterogeneous porous scaffold design for tissue engineering using
  triply periodic minimal surfaces.
\newblock {\em International Journal of Precision Engineering and
  Manufacturing}, 13:527--537, 2012.

\bibitem{yoo2015advanced}
D.-J. Yoo and K.-H. Kim.
\newblock An advanced multi-morphology porous scaffold design method using
  volumetric distance field and beta growth function.
\newblock {\em International Journal of Precision Engineering and
  Manufacturing}, 16:2021--2032, 2015.

\bibitem{yuan2019additive}
L.~Yuan, S.~Ding, and C.~Wen.
\newblock Additive manufacturing technology for porous metal implant
  applications and triple minimal surface structures: A review.
\newblock {\em Bioactive materials}, 4:56--70, 2019.

\end{thebibliography}

\end{document}